\documentclass[11pt,aps,showpacs,nofootinbib,longbibliography]{revtex4-2}

\pdfoutput=1

\usepackage[normalem]{ulem}

\usepackage{comment}
\usepackage[dvipsnames]{xcolor}
\usepackage{amssymb,amsmath,latexsym,mathrsfs}
\usepackage[USenglish,american]{babel}
\usepackage{bm}
\usepackage{amsfonts}
\usepackage{graphicx}
\usepackage{epstopdf}
\usepackage{hyperref}
\usepackage{array}
\usepackage[utf8]{inputenc}
\usepackage{soul}
\usepackage{color}
\usepackage{slashed}
\usepackage{csquotes}
\usepackage[T1]{fontenc}
\usepackage{braket}

\usepackage{footnote}

\everymath{\displaystyle} 

\usepackage{soul}

\newcommand{\beq}{\begin{equation}}
\newcommand{\eeq}{\end{equation}}
\newcommand{\beqa}{\begin{eqnarray}}
\newcommand{\eeqa}{\end{eqnarray}}
\newcommand{\bea}{\begin{eqnarray}}
\newcommand{\eea}{\end{eqnarray}}

\begin{document}

%\title{CQC vs. Stochastic: the performance in computing backreaction from semiclassical simulations}

\title{Semiclassical Backreaction: A Qualitative Assessment}

\author{Fabio van Dissel} 
\email{\baselineskip 11pt fvdissel@ifae.es}
\affiliation{ Institut de F\'{\i}sica d'Altes Energies (IFAE)\\ 
 The Barcelona Institute of  Science and Technology (BIST)\\
 Campus UAB, 08193 Bellaterra (Barcelona) Spain
 }

\author{George Zahariade} 
\email{\baselineskip 11pt george.zahariade@uj.edu.pl}
\affiliation{Instytut Fizyki Teoretycznej, Uniwersytet Jagiello\'nski, Lojasiewicza 11, 30-348 Krak\'ow, Poland
 }

\begin{abstract}

%In this short paper, we investigate the performance of two popular methods of computing semiclassical corrections to classical equations of motion against the full quantum mechanical evolution of a simple toy model of two coupled harmonic oscillators. On the one hand, we use a mode-function expansion in which a real quantum degree of freedom is evolved as a complex classical degree of freedom in which the expectation values of the quantum operator can readily be expressed. This expectation value then corrects the evolution of the classical background. On the other hand, we use a stochastic approach (often referred to as the truncated Wigner approach in the literature), in which we sample many initial conditions from the Wigner functional of the initial state of the system. Each I.C. is then evolved classically. In this method expectation values appear as averages over all the different simulations. 

The backreaction of quantum degrees of freedom on classical backgrounds is a poorly understood topic in theoretical physics. Most often it is treated within the {\it semiclassical approximation} with the help of various {\it ad hoc} prescriptions accounting for the effect of quantum excitations on the dynamics of the background. We focus on two popular ones: (i) the {\it mean-field} approximation whereby quantum degrees of freedom couple to the classical background via their quantum expectation values; (ii) the (stochastic) {\it Truncated Wigner} method whereby the fully coupled system is evolved using classical equations of motion for various randomly sampled initial conditions of the quantum degree of freedom, and a statistical average is performed {\it a posteriori}. We evaluate the performance of each method in a simple toy model against a fully quantum mechanical treatment, and identify its regime of validity. We interpret the results in terms of quantum entanglement and loss of classicality of the background.

\end{abstract}

\maketitle

\newpage 

\tableofcontents

\newpage

\section{Introduction}
\label{sec:intro}

Since the dawn of quantum mechanics, understanding the dynamics of quantum variables evolving in classical background fields has played a central role in the development of fundamental physics. In fact some of the foundational experiments of quantum mechanics, such as the Stern-Gerlach experiment \cite{Gerlach_Stern_1922} for instance, rely on the existence of a classical external electromagnetic field which interacts with the quantum system. This {\it semiclassical} point of view, where part of the system is treated classically while another part is treated quantum mechanically, is well-suited for situations where either (i) the external field is actively maintained by an operator, or (ii) the field is so strong (in a sense that can be made precise) that any fluctuations due to its interactions with the quantum degrees of freedom are negligible. In both of these cases, the dynamics of the background is unaffected by the presence of the quantum degrees of freedom, and the external field obeys classical equations of motion.

This approach has been tremendously successful in explaining the splitting of atomic energy levels in the presence of external electromagnetic fields \cite{ZEEMAN_1897, Stark_1914} (Zeeman and Stark effects), and is routinely used in quantum optics~\cite{Mandel_Wolf_1995,Grynberg_Aspect_Fabre_2010}, where it allows for the accurate description of laser-matter interaction, and of many features of the photoelectric effect.\footnote{Historically, the photoelectric effect was understood as a {\it smoking gun} for the quantization of the electromagnetic field and the existence of the photon, however it turns out that its most salient features can be understood as arising from the interaction of quantum  matter with a classical electromagnetic field (see e.g. Chapter 9 or Ref.~\cite{Mandel_Wolf_1995} or Chapter 2E of Ref.~\cite{Grynberg_Aspect_Fabre_2010}).} Within the framework of quantum field theory, it also predicts the creation of electron positron pairs in a strong enough electric field, phenomenon known as {\it Schwinger pair production} \cite{PhysRev.82.664}.

The exact same paradigm applies to situations where the external field is no longer electromagnetic in nature. This is for instance the case in condensed matter when studying quantum fluctuations on top of Bose-Einstein condensates~\cite{Pethick_Smith_2008}, and more generally in field theory where the quantum stability of long-lived non-perturbative classical solutions such as solitons and topological defects~\cite{rajaraman1982solitons}, oscillons, and q-balls~\cite{Coleman:1985ki} has been studied~\cite{Dashen:1974ci,Dashen:1974cj,Dashen:1975hd,Hertzberg_2010, Tranberg_2014}. But the external field can also be gravitational in nature. This is covered in detail within the framework of quantum field theory in curved spacetime~\cite{Birrell:1982ix} and leads to important phenomena such as cosmological particle production \cite{Ford:1986sy,Ford_2021} or Hawking radiation \cite{Hawking:1974rv,Hawking:1975vcx}.

The latter example is particularly important, since it suggests that black holes radiate thermally and thus, by energy conservation, lose mass and evaporate. Such a situation lies precisely outside the regime of validity of the semiclassical regime of quantum mechanics described above since the classical background (the metric around a Schwarzschild black hole for instance) must be sensitive to the production of Hawking quanta. This {\it backreaction} of quantum radiation on the classical background that has generated it, lacks a satisfying description since a full theory of quantum gravity is unavailable for the time being. One possible way of taking quantum backreaction into account is to replace the energy-momentum tensor on the right hand side of the Einstein equations with its quantum expectation value taken in a properly chosen quantum state. This of course comes with its own difficulties because surely such a {\it mean field} approximation would only be valid as long as the quantum state stays sufficiently regular and doesn't develop a spatially multi-modal character (the analog of a double-humped spatial probability density in quantum mechanics).\footnote{Not to mention the many subtleties relating to the renormalization of the usual quantum field theoretic divergences accompanying correlation functions in the coincident limit~\cite{Birrell:1982ix}.}

While the mean-field approximation can be applied in any context (and not just to semiclassical gravity) by replacing any ocurrences of the quantum degrees of freedom in the classical equation of motion for the background with their quantum expectation values, it is just one of the possibilities. Another is to treat the fully coupled system classically but to sample the initial conditions for the quantum degree of freedom from a properly chosen probability distribution (usually specified by the relevant quantum state). Of course the initial condition for the background would be uniquely defined since it is supposed to be classical. Each initial condition will yield a particular realization of the dynamics and one could expect that averaging over a large number of realizations would produce the quantum backreacted dynamics of the classical background. This {\it stochastic} method has been used in the context of early universe cosmology and (p)reheating in particular \cite{Braden_2019, Figueroa_2021}

Of course there are various other ways of taking backreaction into account, with the help of Langevin or Fokker-Planck equations~\cite{Ford:1988zz,Hu:1994ep,Klauder:1983sp}, by integrating out quantum fluctuations and using an effective action to determine the background dynamics~\cite{PhysRev.127.965,Donoghue:1994dn}, or by using open quantum field theory methods akin to the so-called {\it in-in} or Schwinger-Kheldysh formalism~\cite{Calzetta:1986ey,Hu:1994ep,Hu:2008rga}. However all of these methods are quite unwieldy and often times lead to computations that, although in principle doable to arbitrary precision (see for instance Ref.~\cite{PhysRevD.109.085019} where the reflection coefficient on a so-called reflectionless kink is computed in a fully quantum field theoretic manner and backreaction is included), are difficult to perform in practice. Moreover, to our knowledge none of the above semiclassical backreaction methods has been checked against a fully quantum mechanical treatment of the coupled system (either because of the intrinsic complexity of the task or because such a treatment is unavailable as in the case of gravity).

In this work we will be focusing on the above mean field and stochastic methods only. The goal is to gauge their performance in a physically motivated toy model where a fully quantum mechanical numerical treatment is within reach. In Section~\ref{sec:motivation} we will give some background and motivate our approach in the context of previous work on the so-called {\it quantum break time} of a system \cite{Dvali_2017, Dvali_2018, Dvali_20182,Dvali_2022np, michel2023timescalesquantumbreaking, Berezhiani_2022}. In Section~\ref{sec:model} we will introduce the cosmologically inspired toy model that we will be considering (two simple harmonic oscillators coupled via a bi-quadratic interaction) and set up the two semiclassical backreaction methods whose performance will be assessed. In Section~\ref{sec:results} we will showcase our results, outlining the regions of parameter space where they should be trusted and for how long. Finally, we will end in Section~\ref{sec:disc} with a broad discussion of the role that classical instabilities of the background and quantum entanglement play in the breakdown of the semiclassical approximations used widely in the literature.

\section{Motivation}
\label{sec:motivation}
The standard way of performing computations in quantum field theories is to look for an extremum of the classical action, the classical background, and expand the path integral around it, using perturbation theory in the coupling parameter (or equivalently in $\hbar$
). If the background is static, this procedure is well defined to any order and is particularly well suited for analyzing scattering problems. However, if it has some nontrivial time dependence, there is only an adiabatic notion of vacuum for the quantum fluctuations living on top of it. %\gz{[Before you wrote "for the theory" which is imprecise.]} 
This means that the time-varying background will generically create particles ``out of nothing.'' Since energy conservation holds, it is expected that the creation of these particles will affect the classical background, entangling it with the first-order perturbations appearing in the saddle point approximation. There is at this point no clear consensus in the literature on how to deal with the question of backreaction in such situations and the present work aims to shed some light on the issue.

To highlight some of the important ideas we will take the following Lagrangian describing the dynamics of two real scalar fields, $\phi$ and $\chi$, in flat 4 dimensional spacetime, and interacting via a so-called {\it portal} interaction, as a starting point,\footnote{We use the mostly minus signature for the Minkowski metric.}
\beq
\mathcal{L} = \frac{1}{2}\partial_\mu \phi \partial^\mu \phi + \frac{1}{2} \partial_\mu \chi \partial^\mu \chi - \frac{1}{2}\omega^2\phi^2 - \frac{1}{2} \nu^2 \chi^2 - \frac{\lambda}{2} \phi^2 \chi^2  ,
\label{eq:lagrangianpre}
\eeq
where we work in units where $c=1$ (but keep explicit factors of $\hbar$ for the time being), $\omega$, $\nu$ are two angular frequencies (corresponding to the mass parameters or inverse Compton wavelengths of the two fields), and $\lambda$ is the coupling constant. %\gz{[Now I'm happy with the units.]}

This type of lagrangian is of particular interest in early universe cosmology as it represents a popular toy model to study (p)reheating after inflation, although here we will neglect the effects of expansion and work in Minkowski spacetime. To study this model, standard lore tells us to start from an extremum of the classical action. In the context of reheating, one often considers one of the fields (by convention $\phi$) to play the role of the inflaton, while the other field ($\chi$) is simply a spectator. In particular, a natural solution for the physics at the end of inflation is given by
\beq
\phi(t,\vec{x}) = \phi(t) = \phi_0 \cos(\omega t) \quad{\rm and}\quad\chi(t,\vec{x}) = 0.
\label{eq:sadddlepre}
\eeq
Equation~\eqref{eq:sadddlepre} represents an analytic solution of the classical equations of motion, valid for all times. As mentioned above, this is only true in the classical limit of $\hbar \rightarrow 0$, and the quantum mechanical fluctuations of the fields will alter these dynamics. 
To understand how, and to make the connection between classical and quantum physics, we then ask the question of whether there exists a quantum mechanical state (in the Heisenberg picture) for which the 1-point functions $\langle\hat{\phi}(t,\vec{x})\rangle$ and $\langle \hat{\chi}(t,\vec{x}) \rangle$ obey Eqs.~\eqref{eq:sadddlepre}, and which is, at least approximately, Gaussian and ``well localized'' in field phase space. In other words we require (i) that the higher order correlation functions can be related to the 2-point ones via Wick's formula, and (ii) that these 2-point functions approximately factorize up to vacuum contributions \cite{Glauberfact} e.g. $\langle\hat{\phi}(x)\hat{\phi}(x')\rangle\approx\langle\hat{\phi}(x)\rangle\langle\hat{\phi}(x')\rangle+\langle 0|\hat{\phi}(x)\hat{\phi}(x')|0\rangle$.
%and for which the variance for the higher point functions is minimal so that $\langle \hat{x}^n \rangle \approx \langle \hat{x}\rangle^n$.
In other words, are there states in quantum mechanics that can (almost) be described by classical physics?

Of course, such states exist and are the well-known coherent states of quantum mechanics~\cite{glaubercoh}. Formally, a  coherent state for the mode $\vec{k}$ is an eigenstate of the corresponding annihilation operator $\hat{a}_{\vec{k}}$. It is a state $\ket{\mathcal{N}_{\vec{k}}}$, populated by an on average constant number of particles $\langle \hat{N}_{\vec{k}} \rangle = \bra{\mathcal{N}_{\vec{k}}} \hat{a}^\dagger_{\vec{k}} \hat{a}_{\vec{k}} \ket{\mathcal{N}_{\vec{k}}}=\mathcal{N}_{\vec{k}}$. (In particular, the state with no particles in the mode $\vec{k}$, $\ket{0_{\vec{k}}}$, is a coherent state.) A field coherent state is obtained by taking the tensor product of mode coherent states. The coherent state corresponding to Eqs.~\eqref{eq:sadddlepre} is obtained by populating the homogeneous mode of the $\phi$ field with $\mathcal{N} = \frac{\phi_0^2}{2 \omega^2 \hbar}$ particles per (Compton) volume $\omega^{-3}$ while keeping all other modes (including those of the field $\chi$) empty. Although we can formally construct this state in the Heisenberg picture, there is no guarantee that the desirable properties mentioned above survive during time evolution. In fact, for $\lambda \neq 0$, the particles in the coherent state will scatter, get entangled\footnote{By this we simply mean that the quantum state of the coupled system will no longer be {\it separable} i.e. factorizable as a tensor product of individual mode states.} with each other, and lose their original coherence. After some time the 1-point function will no longer correspond to the solution of the classical equations of motion. At that time a full quantum treatment is required to compute observables, as higher order correlation functions cease to be computable from the classically obtained solutions. This time has been referred to as the {\it quantum break time} in the literature \cite{Dvali_2017}. Interestingly, the classical limit of $\hbar \rightarrow 0$ is equivalent to the double scaling of the dimensionless coupling $\alpha = \hbar \lambda \rightarrow 0$ and the average occupancy of the coherent state $\mathcal{N} \rightarrow \infty$, with the so-called {\it collective coupling} $\alpha \mathcal{N}$ kept finite (which is possible since it just contains classical variables). It is therefore good to keep in mind that many coherent particles correspond to weak coupling, which in turn corresponds to the classical limit.

For any finite value of $\hbar$ we expect a finite value of the quantum break time, which was argued to scale as $t_q \sim 1/(\omega \alpha^2 \mathcal{N})$ in \cite{Dvali_2017}. Since a full quantum treatment is usually difficult,\footnote{Although some attempts have been made to study this problem by explicitly constructing high occupancy coherent states in an {\it interacting} quantum field theoretic setting \cite{Berezhiani_2021, Berezhiani_2022, Ilderton_2018, copinger2024paircreationbackreactionresummation, berezhiani2023perturbativeconstructioncoherentstates}.} simply because of the dimensionality of the problem, various semiclassical approximations have been proposed to study quantum corrections in the limit of large occupation number $\mathcal{N}$ and be able to evolve the problem beyond the quantum break time. Generically, all of these methods assume $\phi$ to be classical and $\chi$ to be quantum, and prescribe the way in which these two fundamentally different degrees of freedom interact with one another.
However, since in most cases we don't have an exact quantum field theoretic computation to compare to, the validity of these approximations is untested. As mentioned in the introduction, we are interested in comparing two such popular methods to a full quantum mechanical calculation. We are thus forced to limit ourselves to a toy model of two coupled quantum harmonic oscillators. This can be thought of as the field theory model of Eq.~\eqref{eq:lagrangianpre} restricted to the homogeneous mode of the field $\phi$ and just one of the {\it real} modes of the field $\chi$. 
We are thus interested in computing the effect that the production of quanta in a $\chi$ field mode has on the dynamics of the homogeneous mode of the classical field $\phi$. In other words we are interested in the semiclassical backreaction of $\chi$ on $\phi$ in a minisuperspace approximation. In the next section, we introduce the toy model that we will consider as well as the two semiclassical approximation schemes. From now on we will focus exclusively on this simplified system but we will also be touching upon concepts that are important to the more general case outlined here in this section.

\section{Model}
\label{sec:model}
In what follows we will be working with a quantum mechanical system consisting of two simple harmonic oscillators coupled via a bi-quadratic interaction term. The dynamics of the model will thus be described by the Hamiltonian 
\beq
H = -\frac{\hbar^2}{2 m}\partial_x^2 -\frac{\hbar^2}{2 M}\partial_y^2 + \frac{1}{2} m \omega^2 x^2 + \frac{1}{2} M \Omega^2 y^2 + \frac{\lambda}{2} x^2 y^2,
\label{eq:hamiltoniantoy}
\eeq
where $m$, $M$, $\omega$, $\Omega$ are the masses and  angular frequencies of the oscillators, and $\lambda$ is the coupling strength.

In analogy with the model of Sec.~\ref{sec:motivation}, the (Heisenberg picture) state of the system, defined at the initial time $t=0$ and denoted by $\ket{x_0,0}$, should be prepared in such a way that the $x$ variable is in a coherent state with large occupation number (or in other words $\bra{x_0,0}\hat{x}(t=0)\ket{x_0,0}=x_0\gg \sqrt{\hbar/2m\omega}$), thus playing the role of the classical homogeneous background $\phi$, while $y$ is in its ground state, thus playing the role of one of the $\chi_{\vec{k}}$. This can be achieved in the following way: (i) we start with the free Hamiltonian and both $x$ and $y$ in their respective ground states; (ii) we then adiabatically displace the potential for $x$ to a position $x_0$, which has the effect of changing the ground state for $x$ into a coherent state; (iii) we adiabatically turn on the interaction term which mildly entangles the two degrees of freedom with one another. At this point the system is in the interacting vacuum of the potential $\frac{1}{2} m \omega^2 (x-x_0)^2 + \frac{1}{2} M \Omega^2 y^2 + \frac{\lambda}{2} x^2 y^2$ and thus has no dynamics. The sudden recentering $\frac{1}{2}m\omega^2(x-x_0)^2\to\frac{1}{2}m\omega^2x^2$ at $t=0$ starts the evolution of the system with dynamics given by~\eqref{eq:hamiltoniantoy}.%\gz{[Too much detail?]} 

We will be concerned with various ways of numerically finding the dynamics of $\langle \hat{x}(t)\rangle=\bra{x_0,0} \hat{x}(t)\ket{x_0,0}$ in this setup. On the one hand, because our toy model is simple enough, it is feasible to solve the full Schr\"odinger equation numerically, thus obtaining a reliable benchmark that we can compare other methods to. On the other hand, we can treat $\langle \hat{x}(t)\rangle$ as a classical variable whose classical dynamics incorporate the semiclassical backreaction of the quantum variable $y$. As mentioned before, this can be achieved in various ways but we will focus on two specific ones.
As we will show, the quantum coupling will in some cases cause large entanglement of the variables, making any attempt at understanding the dynamics of $\langle \hat{x}(t)\rangle$ semiclassically, fruitless.

Let us pause for a moment to define what we mean by entanglement here and more generally throughout this paper. Generally, an entangled quantum state is defined in opposition to a product (or separable) state. Thus, in the case of our toy model, an entangled state will be a state $\ket{\psi}$ that cannot be written as a tensor product $\ket{\rho}_x\otimes\ket{\sigma}_y$ of an $x$ state and a $y$ state. There are many measures of entanglement including {\it purity}, {\it entanglement entropy}, and more generally {\it Renyi entropies}~\cite{Horodecki:2009zz}, but here we will mostly adopt a less sophisticated approach. We will call a state that is separable up to terms of order $\lambda$ a {\it weakly entangled} state, and conversely, a state that does not satisfy this property will be called {\it strongly entangled}. At the end of Sec.~\ref{sec:results} we will check that this naive approach agrees with more quantitative measures of quantum entanglement. 

Before going any further it is useful to rescale the variables and parameters so that we are strictly working with dimensionless quantities. This will also help in making the connection with the general quantum field theoretic case. 
We start by defining $\tilde{x} = x\sqrt{m\omega/\hbar}$ and $\tilde{y} = y \sqrt{M\Omega/\hbar}$. 
The Hamiltonian of Eq.~\eqref{eq:hamiltoniantoy} can then be recast as
%
% \beq
% H = \hbar \omega \left(-\frac{1}{2}\partial^2_{\tilde{x}}+ \frac{1}{2} \tilde{x}^2\right) + \hbar \Omega \left(-\frac{1}{2}\partial^2_{\tilde{y}}+ \frac{1}{2} \tilde{y}^2\right) + \frac{ \hbar^2 \lambda}{2 m \omega M\Omega} \tilde{x}^2 \tilde{y}^2.
% \label{eq:rescaledH1}
% \eeq
%
%An illuminating way to rewrite this is
%
\beq
\tilde{H} =-\frac{1}{2}\partial^2_{\tilde{x}}+ \frac{1}{2} \tilde{x}^2 + \frac{\Omega}{\omega} \left[-\frac{1}{2}\partial^2_{\tilde{y}}+ \frac{1}{2}\left(1+\tilde{\lambda}\tilde{x}^2\right) \tilde{y}^2\right],
\label{eq:rescaledH2}
\eeq
where we have introduced the dimensionless Hamiltonian $\tilde{H}=H/\hbar\omega$ and dimensionless coupling $\tilde{\lambda} = \frac{\hbar \lambda}{m \omega M \Omega^2}$.
Since the average occupation number in the $x$ degree of freedom for the state $\ket{x_0,0}$ is given by $N = \frac{m \omega x_0^2}{2 \hbar}=\frac{\tilde{x}_0^2}{2}$, we can estimate the relative importance of the different terms in the above expression. The term outside the parentheses is of order $N$ (in the limit of large occupation number that we are interested in), while the other term is of order $\frac{\Omega}{\omega}(1+\tilde{\lambda}N)^{1/2}$. From this observation, it becomes clear that the limit $N\omega/\Omega \rightarrow \infty$ with $N\tilde{\lambda}=\frac{\lambda x_0^2}{2M\Omega^2}$ kept fixed corresponds to a classical decoupling limit where the $x$ variable's dynamics are those of a pure coherent state and are not influenced by the presence of $y$. In a semiclassical treatment, where the full quantum dynamics of $x$ are reduced to those of its average expectation value, this will correspond to the {\it vanishing quantum backreaction limit}. Notice that for a given ratio $\omega/\Omega$, this can be thought of as the limit $N \rightarrow \infty$, $\tilde{\lambda}\rightarrow 0$, keeping the {\it collective coupling} $\tilde{\lambda}N$ fixed, which is manifestly the direct analog of the double scaling we already encountered in Sec.~\ref{sec:motivation}. %In what follows we fix this ratio at $t = 0$ to be some fixed value and focus on the effect of the coupling $\lambda$. 
This double scaling behavior is automatically satisfied in the limit $\hbar\rightarrow0$.

For numerical convenience, we rescale $y$ one last time by defining $\tilde{\tilde{y}}=\tilde{y}\sqrt{\omega/\Omega}=y\sqrt{M\omega/\hbar}$, and $\tilde{\tilde{\lambda}}=\tilde{\lambda}\left(\Omega/\omega\right)^2=\frac{\hbar\lambda}{mM\omega^3}$, and, getting rid of the tildes and double tildes, we obtain the form of the Hamiltonian we will be working with in the rest of the paper,
\beq
H = -\frac{1}{2}\partial^2_x+ \frac{1}{2} x^2 -\frac{1}{2}\partial^2_y+ \frac{1}{2} \left[\left(\Omega/\omega\right)^2 + \lambda x^2\right] y^2.
\label{eq:dimensionlessham}
\eeq
We now turn to the numerical study of the dynamics given by this quantum Hamiltonian.

\subsection{Numerical methods}
\label{sec:sims}

In what follows we will compare the fully quantum mechanical evolution of $\langle \hat{x}(t) \rangle$ obtained from a numerical solution of the Schr\"odinger equation with Hamiltonian~\eqref{eq:dimensionlessham}, to the one predicted by two semiclassical approximation schemes for including the effects of the quantum backreaction of $y$ on the (classical) dynamics of $\langle \hat{x}(t)\rangle$. We will refer to these two methods as the {\it Mean Field} (MF) and {\it Truncated Wigner} (TW) methods.\footnote{Note that what we call MF method is known in the literature by many names: Hartree, Hartree-Fock or simply {\it semiclassical approximation} (especially in a semiclassical gravity context). Likewise the TW method is sometimes known as the {\it stochastic} or {\it statistical} method.}

\subsubsection{Quantum Mechanics}

As discussed at the beginning of the section, the Schr\"odinger picture state at $t=0$ is prepared in such a way that its wavefunction can be written (to first-order in $\lambda$) as
\beq
\psi_0(x, y) = \mathcal{N}_0 e^{-(x-x_0)^2/2} e^{-\left(\left(\Omega/\omega\right)^2 + \lambda x_0^2\right)^{1/2} y^2/2},
\label{eq:wavefunctioninit}
\eeq
where $\mathcal{N}_0$ is a normalization factor.
This corresponds to the tensor product of a coherent state with a large occupation number, $N=x_0^2/2$, for the $x$ degree of freedom, with the zeroth order adiabatic vacuum of the $y$ degree of freedom (whose width is corrected by the interaction term). We stress that this is an approximate expression valid up to order $\lambda$ in perturbation theory. In particular higher order corrections may reveal non-Gaussianity and entanglement. 

Starting from these initial conditions we will solve the Schr\"odinger equation numerically using a second-order symplectic integrator (see Appendix~\ref{app:numerics}). At any point in time we can then output $\langle \hat{x}(t) \rangle$ and use this to assess the performance of the semiclassical approximation schemes. We outline these schemes next.

\subsubsection{Semiclassical approximations}
The starting point of any semiclassical method is the classical equations of motion stemming from Hamiltonian~\eqref{eq:dimensionlessham},
\beqa
&&\Ddot{x} + \left(1 + \lambda y^2\right)x = 0,
\label{eq:class}\\
&&\Ddot{y} + \left[\left(\Omega/\omega\right)^2 + \lambda x^2 \right]y = 0.
\label{eq:quantum}
\eeqa
As mentioned in Sec.~\ref{sec:motivation} and at the beginning of Sec.~\ref{sec:model}, we have prepared our quantum system in such a way that its dynamics yield $\langle \hat{x}(t) \rangle = x_0 \cos t$ and $\langle \hat{y}(t) \rangle = 0$ i.e. such that the 1-point functions obey the above classical equations of motion with initial conditions $x(0)=x_0$ and $y(0)=\dot{x}(0)=\dot{y}(0)=0$. Of course this is only strictly true in the limit $\lambda\to 0$, $x_0\to \infty$ (with $\lambda x_0^2$ constant), or $\hbar\to 0$ in standard units. In fact for large but finite $x_0$ (or small but finite $\lambda$) the full quantum dynamics of the expectation values is expected to agree with the corresponding classical dynamics only until the quantum break time of the system.

The aim of any semiclassical approximation is to continue treating the $x$ degree of freedom classically while relaxing the assumption of classicality for the $y$ degree of freedom. To do this one would need to consistently incorporate the effect of the quantum fluctuations of $y$ on the classical dynamics of $x$ i.e. the quantum backreaction of $y$ on $x$. Without much computational cost, one should thus be able to extend the validity of the resulting, semiclassical, dynamics for $x$ beyond the quantum break time. We will be focusing on two methods implementing this strategy.
\begin{enumerate}
    \item {\it Mean Field approximation}
    
    While keeping $x$ classical we promote $y$ to a quantum operator $\hat{y}$. Working in the Heisenberg picture we can expand it as $\hat{y}(t) = z(t) \hat{a}^\dagger + z^*(t)\hat{a}$, where the creation and annihilation operators are defined with respect to the initial adiabatic ground state such that $\hat{a} \ket{0} = 0$ (see Refs.~\cite{Vachaspati_2018,Vachaspati_2019} for details). The dynamics of the complex function $z(t)$ results from the Heisenberg equations of motion and follows a complexified version of Eq.~\eqref{eq:quantum},
    \beq
        \Ddot{z} + \left[\left(\Omega/\omega\right)^2 + \lambda x(t)^2 \right] z = 0.
        \label{eq:nobrCQC}
    \eeq
    At this point $x(t)$ can be chosen to be $x_0 \cos t$ but in fact it can be any arbitrary background function of time such that $x(0)=x_0$. The appropriate initial conditions for $z(t)$, corresponding to the choice of the adiabatic ground state for $y$, are $z(0) = -i \left(\left(\Omega/\omega\right)^2 + \lambda x_0^2 \right)^{-1/4}$ and $\Dot{z}(0) = \left(\left(\Omega/\omega\right)^2 + \lambda x_0^2 \right)^{1/4}$. As long as the function $x(t)$ is fixed, the solution $z(t)$ of Eq.~\eqref{eq:nobrCQC} is enough to specify the exact quantum dynamics of $y$ i.e. to exactly compute all of its correlation functions. For instance $\langle \hat{y}(t)^2\rangle= |z(t)|^2$.
    However, in this case we are precisely interested in how the background $x$ is corrected, or backreacted upon, by the quantum excitations of $y$. Within the MF approximation, we include this correction in the classical equation of motion for $x$, Eq.~\eqref{eq:class}, by replacing $y^2$ with $\langle \hat{y}^2 \rangle = |z|^2$. The solution, $x(t)$, of the resulting coupled system of equations will thus deviate from the purely classical harmonic and will also appear in  Eq.~\eqref{eq:nobrCQC}. In summary, the set of equations to be solved for in the MF approximation will be
    \beqa
        &&\Ddot{x} + \left(1 + \lambda |z|^2\right)x = 0, \label{eq:classCQC}
        \\
        &&\Ddot{z} + \left[\left(\Omega/\omega\right)^2 + \lambda x^2 \right] z = 0,
        \label{eq:qmCQC}
    \eeqa
    with the adiabatic initial conditions for $z$ and the classical initial conditions for $x(0) = x_0$, $\Dot{x}(0) = 0$. For some practical applications of this technique, including in field theory, see Refs.~\cite{Vachaspati_2018, Vachaspati_2019, Mukhopadhyay_2022, Oll__2019}.
    
    \item {\it Truncated Wigner method}: 
    
    The Wigner function of our initial state~\eqref{eq:wavefunctioninit}  is given by
    \beqa
    W_0(x, y, p_x, p_y) &=&
    \frac{1}{\pi}\int dx' dy' \psi_0^*(x+x',y+y')\psi_0(x-x',y-y')e^{2i(p_x x'+p_y y')}\nonumber\\
    &=&
   \mathcal{N}_0^2 e^{-(x-x_0)^2} e^{-p_x^2} e^{-\left(\left(\Omega/\omega\right)^2 + \lambda x_0^2\right)^{1/2} y^2} e^{-\left(\left(\Omega/\omega\right)^2 + \lambda x_0^2\right)^{-1/2} p_y^2},
    \label{eq:Wignerinit}
    \eeqa
    and can be interpreted as a joint probability distribution for the initial phase space variables of our system of interest. 
    %\gz{[Check. Corrected some factors of 2 in the exponent.]}
    The idea of the TW method is to approximate the time-evolved Wigner function by a truncated sum of $\delta$ functions,
    \beq
    W_t(x, y, p_x, p_y) \approx \frac{1}{N_s} \sum_{i=1}^{N_s} \delta(x - x_i(t)) \delta(p_x - p_{x, i}(t)) \delta(y - y_i(t)) \delta(p_y - p_{y,i}(t)),
    \label{eq:Wignertrunc}
    \eeq
    where $N_s\gg 1$ and different $x_i(t)$, $p_{x,i}(t)=\dot{x}_i(t)$, $y_i(t)$, $p_{y,i}(t)=\dot{y}_i(t)$ are solutions to the classical equations of motion~\eqref{eq:class} and~\eqref{eq:quantum}, with initial conditions $x_i(0)$, $p_{x,i}(0)$, $y_i(0)$, $p_{y,i}(0)$ sampled randomly from the joint probability distribution\footnote{Although the Wigner function does not generically satisfy all the properties required of a {\it bonna fide} probability distribution, in particular positivity, it does satisfy them in the Gaussian case considered here.} of Eq.~\eqref{eq:Wignerinit} \cite{Sinatra_2002}. A nice review of the method is given in \cite{eberhardt2023classicalfieldapproximationultra}. In the following we shall use a variant of this method where all $x_i(t)$, $p_{x,i}(t)$ have definite initial conditions $x_i(0)=x_0$, $p_{x,i}(0)=0$. This corresponds to the assumption of classicality for the $x$ degree of freedom. Likewise the randomness in the initial conditions of the $y$ degree of freedom is supposed to mimic its quantum nature.
    
   The expectation value of any (Weyl ordered) Heisenberg picture observable $\hat{O}(\hat{x}, \hat{p}_x, \hat{y}, \hat{p}_y)$ can then be estimated via
    \beq
        \langle \hat{O}(t) \rangle \approx \frac{1}{N_s} \sum_{i=1}^{N_s} O\left(x_i(t), p_{x,i}(t), y_i(t),p_{y,i}(t)\right),
        \label{eq:Wignerobs}
    \eeq
    or in other words by averaging over the different classical paths. We will be interested in the case $\hat{O}=\hat{x}$ for which we obtain the backreacted dynamics of the classical background 
    \beq
    x(t)=\langle \hat{x}(t)\rangle =\frac{1}{N_s} \sum_{i=1}^{N_s}x_i(t).
    \label{eq:Wignerxobs}
    \eeq
    In summary, this method is implemented by drawing a large number $N_s$ of random initial conditions for $y$ and $p_y$ from the Wigner function~\eqref{eq:Wignerinit} while constraining the initial conditions for $x$ and $p_x$ to be $x_0$ and 0 respectively. Each initial condition is then evolved classically using Eqs. \eqref{eq:class} and \eqref{eq:quantum}, yielding a particular classical realization of the coupled dynamics. Finally, an average over all $N_s$ realizations $x_i(t)$ is performed to obtain a uniquely defined $x(t)$ (see Eq.~\eqref{eq:Wignerxobs}). 
    
    Notice that this method is more computationally intensive that the Mean Field approximation, since $N_s$ usually has to be a large number in order to bring the statistical variance to negligible levels. In this work we typically take $N_s \sim 10^4$. 
    
\end{enumerate}

We will need to compare the results of these two methods to the full quantum mechanical evolution of $\langle \hat{x}(t)\rangle$. Indeed, while we expect the above semiclassical solutions to agree with the exact one for a time longer than the quantum break time, they will eventually deviate from it. We will thus need a measure of ``correctness'' allowing us to decide until what time a given semiclassical method is to be trusted.  

\subsubsection{Measure of correctness and semiclassical break time}
Each of the two semiclassical methods described above gives a value for $x(t)$. We would like to know at what point in time this value deviates significantly from the true value, $\langle \hat{x}(t)\rangle$, obtained by evolving the wavefunction with the Schr\"odinger equation. For this purpose, we first need to introduce an appropriate error function $\delta(t)$. A straightforward way to do this is to use the so-called $L^1$ norm to compute the (relative) integrated difference between semiclassical and full quantum solutions. We thus define
\beq
\delta(t) =  \frac{\int_{0}^t|x(t') - \langle \hat{x}(t')\rangle|dt'}{\int_{0}^t\left(| x(t')| + |\langle \hat{x}(t')\rangle|\right)dt'},
\label{eq:delta2}
\eeq
where $x(t)$ is computed semiclassically either in the MF or TW approximations. The denominator in this definition is simply a normalization factor and is required if we want to use this error function to compare the accuracy of different semiclassical methods across a large region of parameter space. 

We can then define the {\it semiclassical break time} of a specific method to be the time $t$ for which the monotonous function $\delta(t)$ rises above a certain threshold. For most of our results, we shall choose a threshold of $0.05$, but any qualitative conclusions are quite insensitive to this value. In the next section we will estimate the semiclassical break times of the MF and TW methods, referred to as $t_{MF}$ and $t_{TW}$, and compare them to the quantum break time $t_q$ obtained by replacing $x(t)$ in Eq.~\eqref{eq:delta2} with the classical solution $x_{cl}(t)=x_0 \cos t$. We now move on to discussing our results.

\section{Results}
\label{sec:results}

There are three different dimensionless parameters that we can vary in this problem: the coupling $\lambda$, the ratio of frequencies $\Omega/\omega$ (or equivalently the ratio of energies in each mode), and finally $x_0$, which is proportional to the square root of the average occupation number $N$ in the coherent state. As we will see, each of these parameters has a profound effect on the different break times, and on the validity of the semiclassical approximation in general. To understand this point better, it is useful to linearize the classical equations of motion~\eqref{eq:class} and \eqref{eq:quantum} around the solution $x_{cl}(t) = x_0 \cos t$ and $y_{cl}(t) = 0$. We obtain only one non-trivial equation
\beq
\Ddot{\delta y} + \left[ \left(\Omega/\omega\right)^2 + \lambda x_0^2 \cos^2 t\right]\delta y = 0,
\label{eq:Mathieu}
\eeq
which will seem familiar to many readers. It is the famous Mathieu equation, which finds many applications in various areas of physics. The solutions of Eq.~\eqref{eq:Mathieu} can be written as $\delta y(t) \propto p(t) e^{\mu t}$, where $p(t)$ is a periodic function and $\mu$ can in principle be complex (see e.g. \cite{Amin_2014}). Clearly, if $\mu$ has a positive real part, $\delta y$ will grow exponentially signifying that the classical trajectory $(x_{cl}(t),y_{cl}(t))$ is unstable against classical fluctuations and the system becomes parametrically resonant. Whether $\mu$ indeed has a real part is entirely dependent on the three parameters of the problem, $\lambda$, $\Omega/\omega$ and $x_0$. In fact, the three dimensional parameter space $(\lambda,\Omega/\omega,x_0)$ of the Mathieu equation~\eqref{eq:Mathieu} is characterized by the presence of instability bands, centered around the surfaces of equation
\beq
\lambda x_0^2 = 2 \left(n^2 - \left(\Omega/\omega\right)^2\right),
\label{eq:instabbands}
\eeq
where $n$ is a positive integer. When the parameters lie sufficiently close to these surfaces, the system is classically unstable. 
At the quantum level, it has already been conjectured in the literature that the entanglement of the $x$ and $y$ degrees of freedom will grow quickly in the presence of parametric resonance at the classical level \cite{Dvali_2013, Zurek_1994, Bianchi_2018}. We may therefore reasonably expect that this will also lead to qualitatively different behaviors between semiclassical and full quantum dynamics (in other words to shorter semiclassical break times) as our semiclassical approximations struggle to capture purely quantum features such as entanglement. We will show that indeed the instability bands of the Mathieu equation, defined by Eq.~\eqref{eq:instabbands}, will have a profound effect on the semiclassical break times of the two approximation schemes under consideration.

To visualize the break times for the various methods under consideration, we performed a large number of numerical simulations for different combinations of parameters within the  two semiclassical approximation schemes as well as within a full quantum treatment. We obtained independent predictions for the dynamics of the classical $x$ degree of freedom: $x_{MF}(t)$ for the MF method, $x_{TW}(t)$ for the TW method, and $\langle \hat{x}(t)\rangle$ for the exact quantum benchmark. For completeness we also included the classical solution $x_{cl}(t)=x_0 \cos t$.

\begin{figure}[h!]
    \centering
    \includegraphics[width = \textwidth]{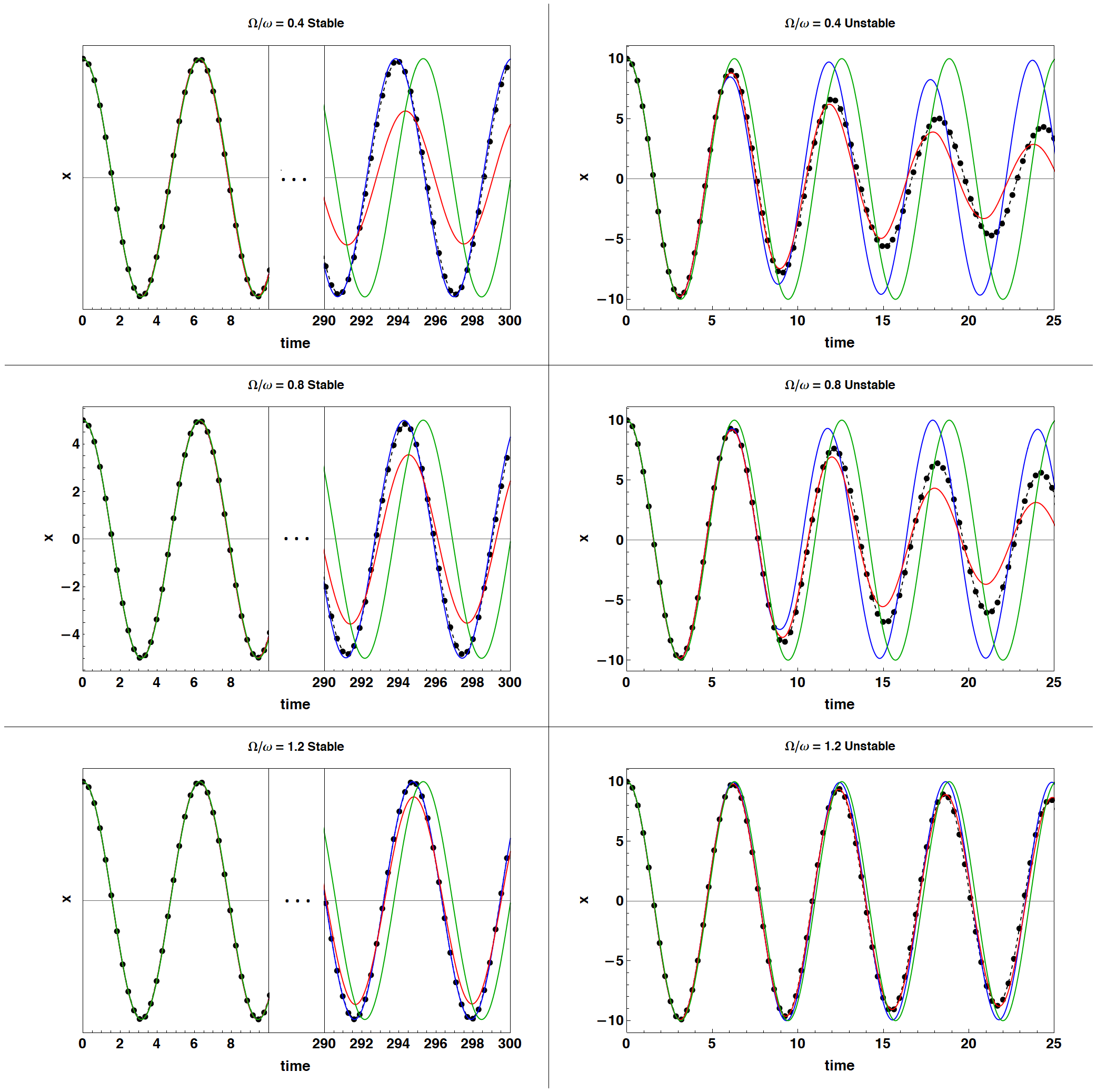}
    \caption{Some typical evolutions of $\langle \hat{x}(t) \rangle$ (black dotted line), $x_{MF}(t)$ (blue line), $x_{TW}(t)$ (red line) and $x_{cl}(t)$ (green line). We illustrate the importance of classical instability by choosing  parameters in such a way that we are either in a completely stable regime (left panels, $\lambda = 0.01$, $x_0 = 5$) or in a regime where the dynamics (at least intermittently) move through an instability band (right panels, $\lambda = 0.1$, $x_0 = 10$). The qualitative difference is visible. In the stable region, we mostly see a small change in the frequency of oscillation (the disagreement between the plots is only visible after some time), whereas in the unstable situation, we see a rapid decay of the amplitude on top of the frequency shift. The different rows correspond to the different choices of $\Omega/\omega$: $0.4$ (upper row), $0.8$ (middle row), and $1.2$ (lower row).} 
    \label{fig:evolutions}
\end{figure}

In Fig.~\ref{fig:evolutions} we show some typical results obtained from our simulations for a range of parameters. Already at the qualitative level, the influence of the instability bands is apparent. 
Parameters have been chosen such that the left panel plots correspond to a stable system while those on the right panel correspond to an unstable, parametrically resonant one, at the classical level. Focusing on the time evolution of $\langle \hat{x}(t)\rangle$, we first notice that its frequency of oscillation gets a small additive correction equal to $\lambda\omega/4\Omega$ (see Appendix~\ref{app:analytics} for details). This has the net effect of slightly shifting the instability bands upwards, their new position still given by Eq.~\eqref{eq:instabbands} but with the substitution $n^2\to n^2\left(1 + \lambda \omega/4\Omega\right)^2$. From now on we will only be referring to these, shifted, instability bands. Notice that, the shift being small, this change in definition has no impact on the classical stability properties of the solutions represented in Fig.~\ref{fig:evolutions}. 

Besides this correction to the oscillation frequency, we also observe a qualitative difference between the stable and unstable cases: while in the stable case the amplitude of oscillation remains more or less constant, the unstable case features a dissipative-like dampening. 
This can be understood heuristically, in a field theoretic language, as follows. In a classically unstable region of parameter space, the particles making up the $x$ mode coherent state rapidly scatter and get entangled with the particles produced in the $y$ mode. The state then ``decoheres''\footnote{Here we simply mean that the state is becoming less like a coherent state. This has nothing to do with the decoherence phenomenon used to explain the classical to quantum transition.} quickly and ceases to mimic a classically oscillating system, thus leading to significant deviations from classical motion. In particular this triggers a dissipative behavior whereby higher and higher moments of the joint probability distribution of $x$ and $y$ are excited and Gaussianity is completely lost~\cite{Bojowald:2005cw,Bojowald:2010qm,Baytas:2018gbu,Baytas:2018ruy,Bojowald_2021,Bojowald_2022}. By contrast, in a stable region of parameter space entanglement still occurs but in a more controlled manner, only leading to the above mentioned multiplicative correction to the oscillation frequency of the $x$ mode coherent state while preserving its Gaussianity. Note that in the stable case where entanglement is expected to be small and controlled, the full quantum dynamics seems to be  better described by the MF method. The opposite seems to be true in the unstable situation where entanglement grows and only the TW method is able to capture dissipative-like effects, although both semiclassical methods fare quite poorly in this case. We will come back to this interesting fact later.

\begin{figure}[h!]
    \centering
    \includegraphics[width = \textwidth]{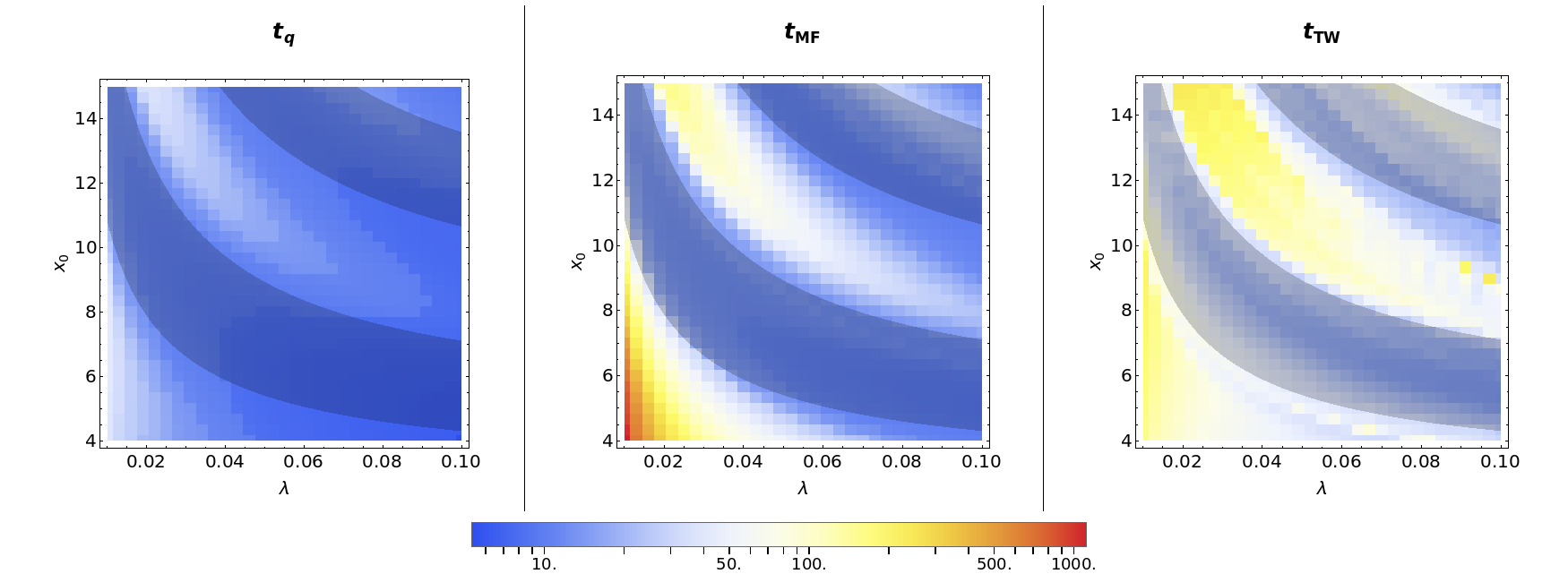}
    \caption{Quantum, and semiclassical break times for the MF and TW methods (as defined via Eq.~\eqref{eq:delta2} and a threshold value of $0.05$) for a region of the $(\lambda,x_0,\Omega/\omega=0.4)$ parameter plane. Redder (bluer) regions correspond to larger (shorter) break times and thus a more (less) accurate approximation of $\langle \hat{x} \rangle$. The shaded regions represent the instability bands of the Mathieu equation (see Eq.~\eqref{eq:instabbands}). Their influence is clear: when the parameters of the system lie in their vicinity, there is a considerable enhancement of quantum entanglement between $x$ and $y$, leading to a breakdown of classicality for $x$.}
    \label{fig:w0d4grid}
\end{figure}

\begin{figure}[h!]
    \centering
    \includegraphics[width = \textwidth]{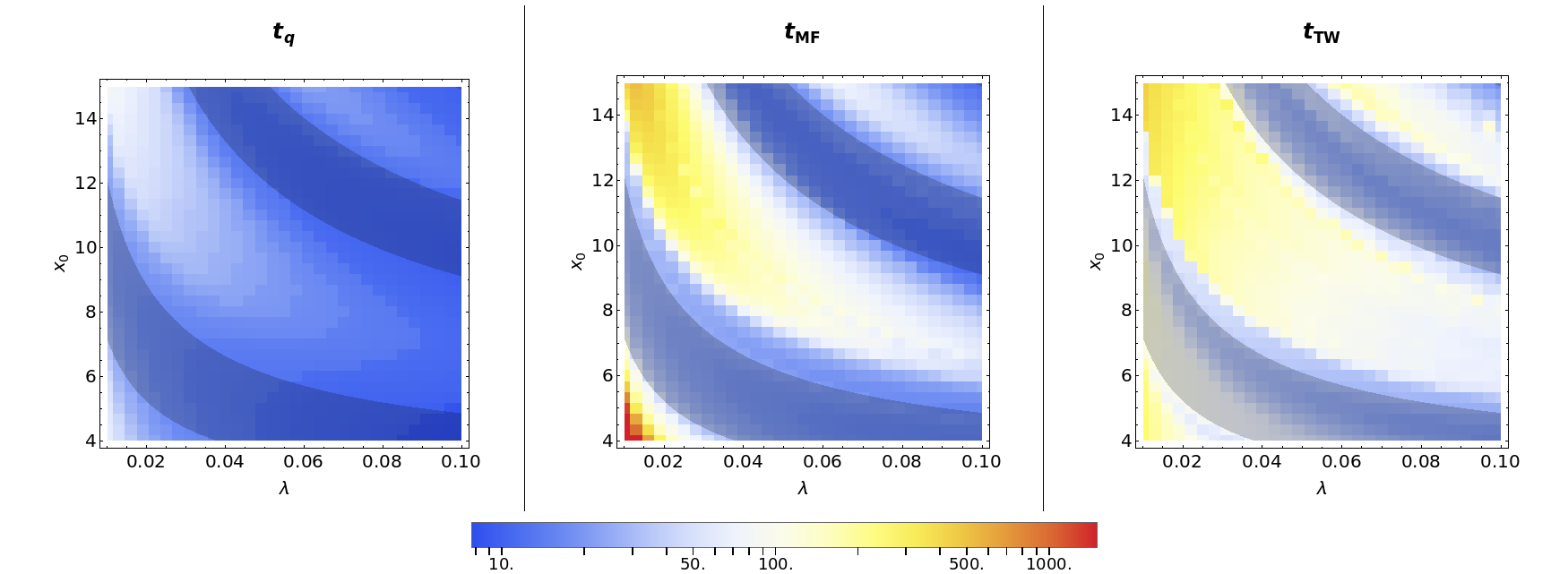}
    \caption{Same plot as Fig.~\ref{fig:w0d4grid} with $\Omega/\omega = 0.8$.}
    \label{fig:w0d8grid}
\end{figure}

\begin{figure}[h!]
    \centering
    \includegraphics[width = \textwidth]{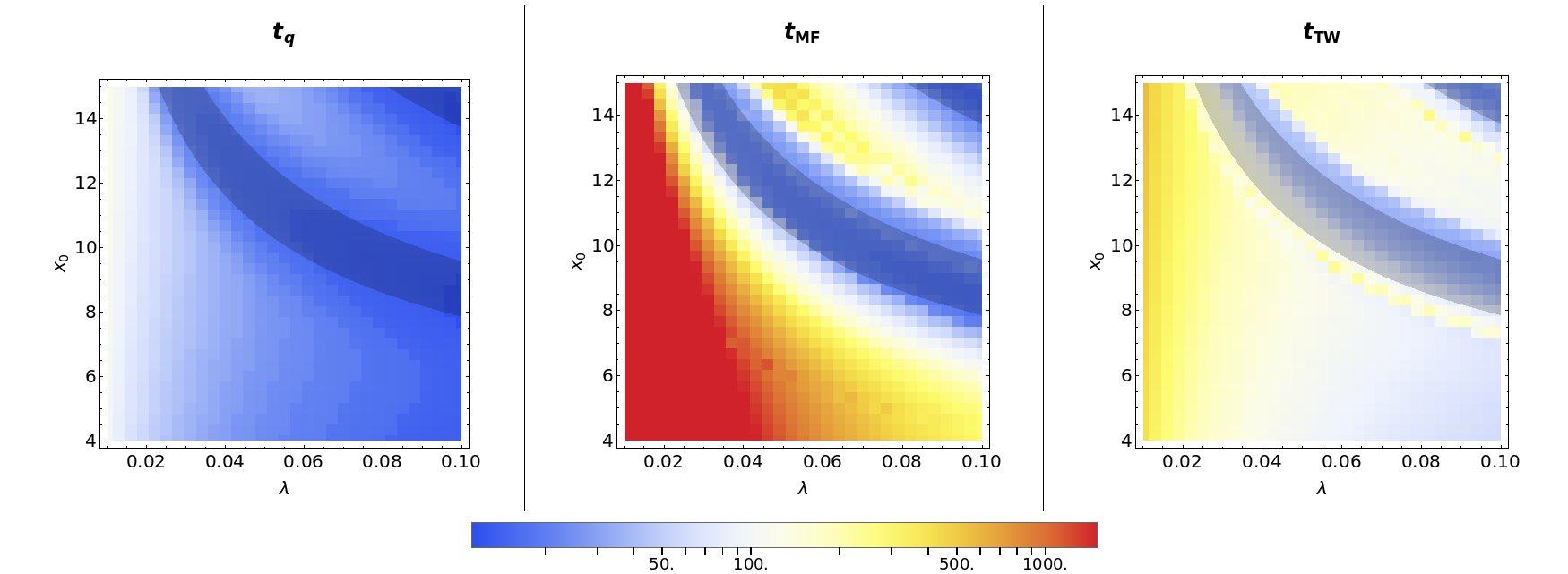}
    \caption{Same plot as Fig.~\ref{fig:w0d4grid} with $\Omega/\omega = 1.2$.}
    \label{fig:w1d2grid}
\end{figure}

The importance of the instability bands is qualitatively clear at the level of the individual simulations, but how does it influence the break times of the various methods for a broad range of parameters? In Figs.~\ref{fig:w0d4grid}, \ref{fig:w0d8grid}, and \ref{fig:w1d2grid} we show the results of a parameter scan of the break times $t_{MF}$, $t_{TW}$, and $t_q$. Recall that these are defined as the times when $\delta(t)$, given by Eq.~\eqref{eq:delta2} with $x(t)$ replaced by $x_{MF}(t)$, $x_{TW}(t)$, and $x_{cl}(t)$ respectively, crosses a predefined threshold value in the simulation. Here we chose $0.05$ as this critical threshold. 
Inspecting Figs.~\ref{fig:w0d4grid}, \ref{fig:w0d8grid} and \ref{fig:w1d2grid}, we see the effect of the classical instability band structure of the Mathieu equation. Around the bands, indicated by the shaded regions, the quantum and semiclassical break times decrease considerably. This supports the understanding that entanglement is at the root of the breakdown of (semi)classicality. The MF and TW methods prolong the agreement between the full quantum dynamics and a semiclassical description of the quantum backreacted dynamics of the $x$ degree of freedom, in certain regions of parameter space considerably. Indeed, generically $t_{MF},\,t_{TW}> t_q$ with a larger time gain far from the classical instability bands. However, both methods perform poorly within the instability bands, even though TW performs marginally better there. Notice also, that because the amplitude of $x(t)$ decreases in the parametrically unstable case, the system dynamically moves out of the instability band in the direction of decreasing $x_0$, which presumably brings the resonant phase to its natural conclusion. Finally, it is noteworthy that break times generally decrease substantially when moving upwards and to the right in parameter space, even outside any instability band, which is correlated with an increase in the collective coupling $\lambda x_0^2/2= \lambda N$. In appendix~\ref{app:analytics} we show perturbatively that this can be understood as an increase in quantum entanglement between $x$ and $y$.

\begin{figure}[h!]
    \centering
    \includegraphics[width =\textwidth]{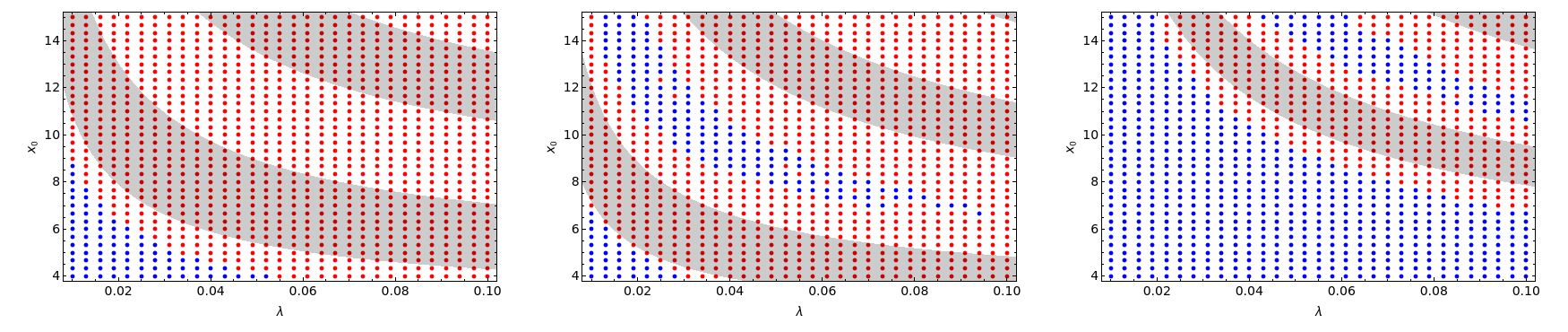}
    \caption{Plot showing the regions in the $(\lambda, x_0)$ parameter plane where $t_{MF} > t_{TW}$ (blue dots) and $t_{MF} < t_{TW}$ (red dots) for different values of the ratio $\Omega/\omega$: $0.4$ (left), $0.8$ center, and $1.2$ (right). The shaded region represent the instability bands. A heuristic picture emerges: far away from the instability bands, the MF method outperforms the TW method. 
    %This suggests the importance of entanglement and hints at several advantages and disadvantages of the two methods. We will discuss these in sec.~\ref{sec:disc}.
    }
    \label{fig:comparison}
\end{figure}

\begin{figure}[h!]
    \centering
    \includegraphics[width =\textwidth]{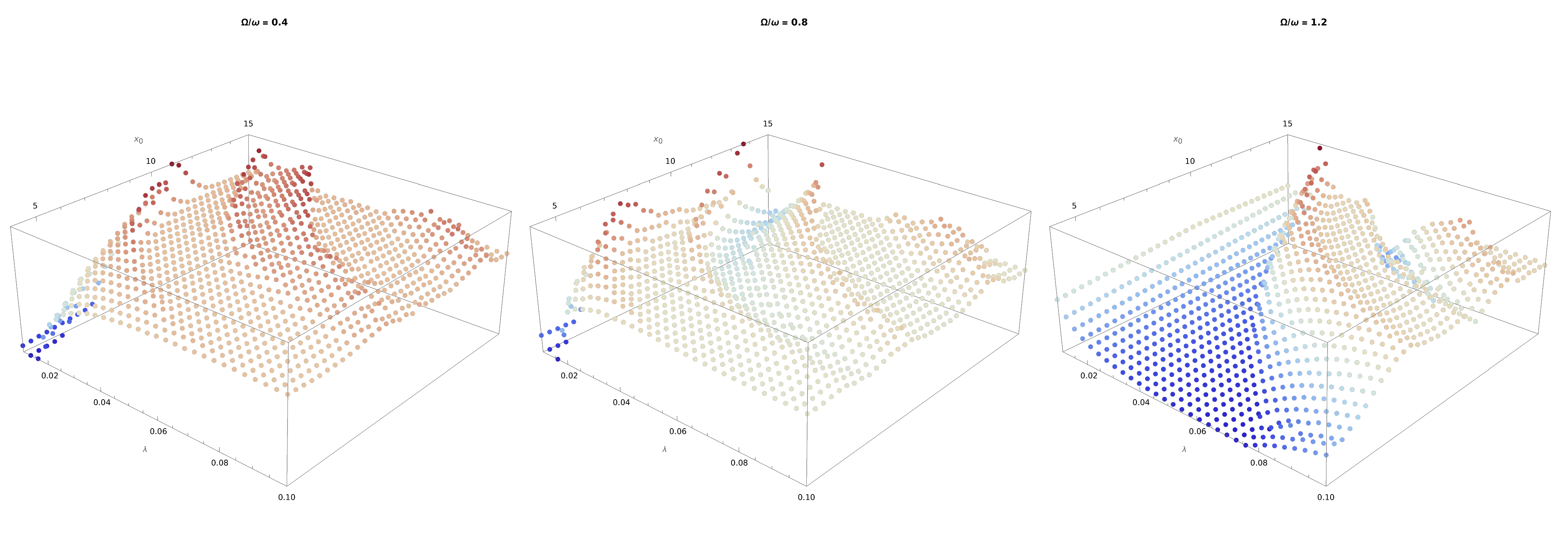}
    \caption{The three dimensional version of Fig.~\ref{fig:comparison}. We plot the difference in break times $t_{TW} - t_{MF}$ as a function of $\lambda$ and $x_0$ for different values of the ratio $\Omega/\omega$: 0.4 (left), 0.8 (center), and $1.2$ (right). Again, the influence of the instability bands is clear.}
    \label{fig:comparison3d}
\end{figure}

It is natural to ask which of the two highlighted semiclassical methods, MF or TW, does a better job at extending the validity of a classical treatment  for the background degree of freedom $x$ in the presence of interactions with the quantum degree of freedom $y$. As seen in Figs.~\ref{fig:w0d4grid},~\ref{fig:w0d8grid}, and~\ref{fig:w1d2grid}, the answer depends on the parameters of the problem. In Fig.~\ref{fig:comparison} we give a representation of the performance of the MF method relative to the TW method, identifying the regions in parameter space where $t_{MF} > t_{TW}$ and vice versa. 
% This gives a sense of the more accurate method for each point in parameter space.
%\begin{figure}[h!]
%    \centering
%    \includegraphics[width =\textwidth]{comparison.png}
%    \caption{The contours in parameter space (left $\Omega/\omega = 0.4$, middle $\Omega/\omega = 0.8$, right $\Omega/\omega = 1.2$) for which $t_{MF} = t_{TW}$, giving a sense of where each method performs better. Within the blue contour $t_{MF} > t_{TW}$ and vice versa. A heuristic picture emerges: far away from the instability bands, MF outperforms TW. This suggests the importance of entanglement and hints at several advantages and disadvantages of the two methods. We will discuss these in sec.~\ref{sec:disc}.}
%    \label{fig:comparison}
%\end{figure}
Fig.~\ref{fig:comparison} suggests that the MF method generally performs better whenever parameters are such that the system is far from any instability band. This highlights the importance of entanglement between the $x$ and $y$ degrees of freedom in evaluating which method to apply: it seems that whenever we are in a situation where entanglement remains under control, the MF method is more adequate in describing the exact quantum dynamics of $\langle \hat{x}(t)\rangle$. However, in situations where $x$ and $y$ become highly entangled, the TW method seems to perform better (although only marginally as seen in Figs.~\ref{fig:evolutions},~\ref{fig:w0d4grid},~\ref{fig:w0d8grid}, and~\ref{fig:w1d2grid}). In Fig.~\ref{fig:comparison3d} we give a sense of the absolute difference in the break times of the two methods by plotting the difference $t_{TW} - t_{MF}$ as a function of $\lambda$ and $x_0$ (for different values of the ratio $\Omega/\omega$).

To summarize our results and to make the connection with quantum entanglement a bit more precise, in Table~\ref{tab:singlemode} we give the break times as well as the entanglement entropy (at some fiducial time) for a few prototypical values of the parameters.
The entanglement entropy of the system is computed as the Von Neumann entropy of the reduced density matrix, $\hat{\rho}_x$,
\beq
\hat{\rho}_x = \sum_m \bra{m} \hat{\rho} \ket{m},
\eeq
where $\ket{m}$ are the eigenstates of the $y$ quantum harmonic oscillator and $\hat{\rho}$ is the full density matrix of the coupled system. As the Hilbert space has infinite dimension we limit the computation to a maximum quantum number $m_{max} = 28$, making sure that the state is contained up to a few parts in $10^2$ in this reduced space. Then the entanglement entropy is given by
\beq
S_e = -\sum_i \lambda_i \log(\lambda_i),
\eeq
where $\lambda_i$ are the eigenvalues of $\hat{\rho}_x$. Heuristically, $S_e$ measures the uncertainty in the ``classical'' $x$ degree of freedom due to the interaction with the ``quantum'' $y$ degree of freedom. Its maximum value is related to the dimensionality of the Hilbert space: in our case $S_{e, max} = \ln(m_{max}) = 3.33$. If and only if there is no entanglement (so that the state is separable) is it exactly $0$. Note that this way of estimating quantum entanglement is just one of many possibilities \cite{Horodecki:2009zz}.

\begin{table}[h!]
\centering
\begin{tabular}{|c|c|c|c|c|c|c|c|}
\hline
\textbf{$\lambda$} & \textbf{$x_0$} & \textbf{$\Omega/\omega$} & Stability & \textbf{$t_q$} & \textbf{$t_{MF}$} & $t_{TW}$ & $S_e$ at $t = 25$ \\ \hline
0.1 & 5& 0.4      & Unstable      & 6      & 9      & 17& 1.86      \\ \hline
0.05 & 8& 0.4      & Unstable      & 9      & 12      & 22& 1.64      \\ \hline
0.03      & 5      & 0.4      & Stable      & 14& 83      & 73      & 0.31      \\ \hline
0.1      & 5      & 0.8      & Unstable      & 10      & 16& 25      & 1.21     \\ \hline
0.04 & 6      & 0.8      & Unstable      & 13      & 20      & 32      & 1.47      \\ \hline
0.05      & 8      & 0.8      & Stable      & 20      & 127& 105      & 0.70     \\ \hline
0.03      & 5      & 1.2      & Stable & 40      & 3304      & 164      & 0.026      \\ \hline
0.04& 6      & 1.2     & Stable & 31      & 1385      & 140      & 0.068    \\ \hline
0.08& 10      & 1.2     & Unstable & 14      & 22      & 34      & 0.13    \\ \hline
\end{tabular}
\caption{A summary of some prototypical results from our numerical computations. The influence of stability on entanglement, and in turn the influence of entanglement on the break times, is evident.}
\label{tab:singlemode}
\end{table}
The results in Table~\ref{tab:singlemode} contain the most salient features of our results. Namely, unstable modes result in larger entanglement, which in turn reduces the break times of the various methods. As suggested earlier, when entanglement is small, one typically has $t_{MF} > t_{TW}$. The quantum break time $t_q$ is always smaller than both $t_{MF}$ and $t_{TW}$, showing that the methods give at least some improvement with respect to a purely classical treatment. Note in particular that for the stable cases of $\Omega/\omega = 1.2$, quantum entanglement is extremely small and $t_{MF}$ is orders of magnitude larger than both $t_q$ and $t_{TW}$.

\begin{figure}[h]
    \centering
    \includegraphics[width=\textwidth]{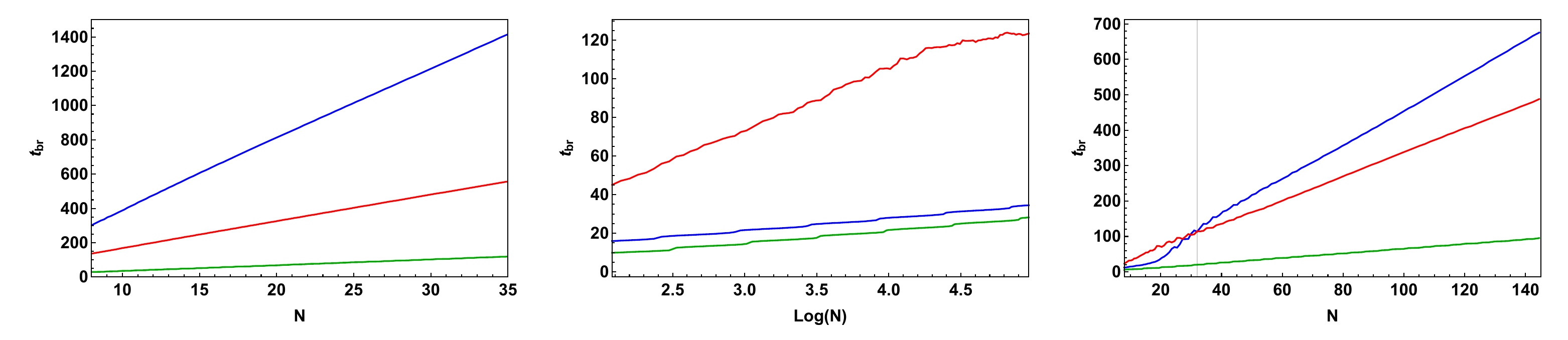}
    \caption{The break times $t_q$ (green), $t_{MF}$ (blue) and $t_{TW}$ (red) plotted against $N=x_0^2/2$ (a proxy for $\hbar^{-1}$) for different values of the collective coupling $\lambda N$ and for $\Omega/\omega=0.8$. The leftmost plot corresponds to $\lambda N=0.15$, and to systems that remain stable all the way through the quantum to classical transition. The middle plot corresponds to $\lambda N=0.6$, and to systems that are unstable. The rightmost plot corresponds to $\lambda N=1.5$, a situation where the system transitions from unstable to stable during the double scaling $\lambda \rightarrow 0$ and $N \rightarrow \infty$. The approximate point of transition is represented by a vertical line on the plot. The expected $N$ and $\log N$ scalings are evident in the stable and unstable situations respectively.}
    \label{fig:tbrvsN}
\end{figure}

Interestingly, and on a slightly different note, we can also use the data from the previous plots to  investigate the classical limit. As alluded to in Sec.~\ref{sec:motivation}, this corresponds to the limit $\hbar \rightarrow 0$, for which $\lambda \rightarrow 0$ and $N\sim x_0^2 \rightarrow \infty$ (while the collective coupling $\lambda N=\lambda x_0^2/2$ is kept constant). For fixed $\lambda x_0^2$, we therefore intuitively expect that the different  break times should increase as $x_0$ increases (and consequently $\lambda$ decreases). In fact, Ref.~\cite{Dvali_2017} uses heuristic arguments to predict break times that scale like $N$. In Fig.~\ref{fig:tbrvsN} we plot $t_q$, $t_{MF}$, and $t_{TW}$ as a function of $x_0$ for various values of the collective coupling $\lambda N$.
 We observe that although the heuristic linear scaling is sometimes correct, the influence of classical instabilities can be considerable, spoiling this prediction. In particular, as has been noted before, classically unstable systems have break times that scale with $\log N$ instead \cite{Dvali_2013, michel2023timescalesquantumbreaking}. For certain values of the collective coupling, one can have a situation where the system exits an instability band during the double scaling limit. There, we observe a transition between the two types of dependence ($\propto N$ and $\propto \log N$), e.g. the rightmost plot of Fig.~\ref{fig:tbrvsN}.

\section{Summary and Discussion}
\label{sec:disc}

In this work we tried to broach the topic of the regimes of validity of different semiclassical approximations. Indeed when classical degrees of freedom are coupled to quantum ones, there is no entirely consistent theory that one can use to make predictions. We therefore only have two options: either (i) find a way of describing the classical degrees of freedom by an appropriate set of quantum degrees of freedom and thus reduce the initial classical-quantum system to purely quantum one; or (ii) come up with approximate methods to evolve the initial system semiclassically i.e. by prescribing how the quantum degrees of freedom should backreact on the dynamics of the classical ones. Since option (i) generically comes with its own set of daunting problems (ambiguities in the quantum description of classical systems, computational complexity, inapplicability to gravitational systems...), progress has mostly relied on the use of various semiclassical approximations. Unfortunately the limitations of such methods have rarely been studied, an important blind spot that we have tried to partially cover. By focusing on a simple toy model involving two quantum, bi-quadratically coupled, harmonic oscillators, whose full quantum dynamics we were able to numerically compute, and taking an appropriate semiclassical limit whereby one (and only one) of the two oscillators becomes effectively classical, we were able to assess the accuracy of two popular semiclassical approximation methods: the Mean Field and Truncated Wigner methods.

We found that the two methods are not nearly equivalent and that their regime of applicability depends on the parameters of the problem. By computing the duration for which a particular semiclassical evolution stays true to the exact quantum evolution (i.e. the semiclassical break time) we noticed that the parametric stability or instability of the associated classical system plays an important role. 

Both semiclassical methods improve upon the purely classical description of the coupled system for all values of the parameters. This is to be expected since they do capture at least some of the quantum backreaction effects that the classical description cannot possibly describe. More interestingly though, for parameters such that the system is not classically parametrically unstable, the MF method outperforms the TW method. This gain in performance seems to be larger for weak collective coupling (product between the coupling and the average occupation number of the coherent state describing the ``classical'' oscillator). On the contrary, for parameters such that the system exhibits classical parametric resonance, the TW method fares better than the MF method. Although the break times are much lower than in the stable case, the TW method is able to capture the dissipative behavior of the system and to track the decay in the amplitude of the classical oscillator. We hypothesize that this behavior is a consequence of the rapid growth of the quantum entanglement between the two oscillators that was pointed out in particular in Ref.~\cite{Bianchi_2018}. There the entanglement entropy was conjectured to grow linearly in time with a coefficient proportional to the sum of the positive Lyapunov exponents of the system.

Based on these observations we can venture a heuristic interpretation of our results. The fact that both methods fare worse in regions of classical instability and large collective coupling highlights the importance of entanglement between the two degrees of freedom. (In appendix \ref{app:analytics} we point out that entanglement is indeed controlled by the magnitude of the collective coupling in the perturbative regime, when the system is classically stable.) It is not entirely surprising then that the TW method is able to track the system better in unstable regions, since the method evolves many more degrees of freedom over its different initializations, and is thus better able to capture the higher order moments of the wavefunction for $y$ that are excited as quantum entanglement grows. Although this improves the ability of the method to capture the effects entanglement, the slow diffusion of the classical paths through phase space always leads to a finite break time. The MF method does not suffer from this problem and can track the evolution of the quantum state for all times, as long as the collective coupling is small and the system is classically stable. %We show this in appendix \ref{app:analytics}. 
Based on these considerations we can write down a heuristic formula for the break times of the two methods:
\beqa
t_{MF} &\propto& \left(\alpha \cdot Entanglement\right)^{-1},\\
\label{eq:tmfheur}
t_{TW} &\propto& \left(\beta \cdot Entanglement + Diffusion\right)^{-1}.
\label{eq:ttwheur}
\eeqa
Note that this should only be taken as a schematic equation, summarizing our results. The break time of the MF method is generally a function of the amount of entanglement between the two systems. While this is also true for TW, its break time also depends on a general diffusive drift between the various classical paths, due to the random nature of the initial conditions. The dependence on entanglement is stronger for MF however and we generally expect $\alpha > \beta$. Thus in unstable regions where entanglement is high, $t_{TW}>t_{MF}$ although both are relatively short, whereas in stable regions, entanglement is suppressed and the presence of the diffusion term forces $t_{TW}<t_{MF}$.

The previous observations have important consequences for some of the major domains of application of these two semiclassical methods. Taking the example of (p)reheating with which we started our discussion in Sec.~\ref{sec:motivation}, we are now able to reinterpret the standard methodology in terms of our analysis. In fact, what is typically done in the literature of preheating, is to compute particle production from parametric resonance of classical fluctuations (typically obtained through cosmological lattice simulations), averaged over a variety of initializations \cite{Figueroa_2021}. In essence this is an application of the Truncated Wigner method to cosmological quantum fluctuations. We now understand why this is the correct choice as parametric resonance and classical instability are central to preheating, and one can expect a fair amount of entanglement between the inflaton and the resonant vacuum modes. The stable analogue of preheating is reheating \cite{Kofman_1997}. Our expectation is that a mean-field approach should be appropriate in this scenario. Note that the period of reheating has traditionally been studied by assuming cubic couplings $\propto \phi \chi^2$, instead of the quartic coupling under consideration here. It is an interesting question to understand how reheating calculations would be altered by using the MF approach. We leave this interesting question for future work. 

Other scenarios that have been treated using either of the approaches in this paper include the quantum decay of breather states (MF method)~\cite{Oll__2019}, kink-antikink scatterings in a quantum vacuum (MF method)~\cite{Mukhopadhyay_2022}, and false vacuum decay (TW method)~\cite{Braden_2019}. It would be interesting to revisit some of these cases and assess their validity given the results in this work.\footnote{Many breather states under consideration in the literature have nonperturbative collective couplings where $\lambda N = \mathcal{O}(1)$. It has been suggested that the threshold $\lambda N = \mathcal{O}(1)$ characterizes a class of macroscopic objects that saturate the entropy bound coming from $2 \to N$ scattering amplitudes of the theory under consideration \cite{Dvali_2021sat}.} It is also important to mention that the quantum stability of soliton, oscillon, and breather backgrounds has been thoroughly investigated in the literature using perturbative methods as for instance in Ref.~\cite{PhysRevD.12.1038}, and more recently in Refs.~\cite{Evslin:2023qbv,Ogundipe:2024chv} where it was argued that semiclassical methods may obscure important details about the quantum state of the background. For instance it was found that the quantum breather decay discussed using semiclassical MF methods in Ref.~\cite{Oll__2019}, resulted from an implicitly non-coherent choice of state for the background, rather than from an intrinsic quantum instability of the breather.

One of the main limitations of this work is its reliance on the assumption that $\langle \hat{x}(t)\rangle$ computed in a full quantum treatment, should always be a good benchmark against which to compare the performance of different semiclassical approximation methods. This is in fact only true as long the wavefunction $\psi(x,y)$ for the coupled system is well localized and more or less monomodal in the $x$ direction. When this ceases to be true, it is clear that the average expectation value cannot possibly represent the dynamics of a classical variable. This introduces a more fundamental upper limit for the duration of validity of {\it any} semiclassical approximation, irrespectively of the associated value of $\delta(t)$ (see Eq.~\eqref{eq:delta2}).   While we have checked that the wavefunction remains approximately Gaussian in the $x$ direction for the duration of our simulation for some parameter choices, we leave a more exhaustive and rigorous study of this issue for future work. Another important limitation is the fact that it is unclear how to extend our results to the many-mode case and eventually to field theory. Appendix~\ref{app:threemodes} gives some partial answers in the case where the ``classical'' mode is coupled to not one but two ``quantum'' modes. Again the presence of unstable modes seems to shorten the validity of the semiclassical methods under consideration but more work is needed in order to really understand the net effect of stable and unstable modes on the background in a many mode scenario. We also plan to tackle this problem in the near future. The results of Appendix~\ref{app:threemodes} do provide some indication that classical instability and entanglement continue to play an important role in field theory, as it does in the quantum mechanical system under consideration here.

The investigations presented in this article do not claim to be the definitive answer to the question: ``Given a specific classical system interacting with a quantum one, which semiclassical method is the most accurate?'' They do however point out some of the subtleties that are often overlooked in the literature and that could in principle have important consequences. In particular the relation between classical parametric resonance and the rapid breakdown of semiclassical treatments may be relevant for early universe cosmology predictions. In future work, we aim to extend the above analysis to systems with different couplings, exhibiting different forms of instabilities (not just parametric resonance) that may be good toy models for gravitational collapse scenarios for instance. This is central to understanding which semiclassical method (if any) is most suitable for the study of black hole evaporation. In particular it might well be that all semiclassical methods break down before a meaningful fraction of the black hole mass has evaporated, thus rendering black hole abundance estimates moot~\cite{dvali2024memoryburdeneffectblack}.  We also plan on assessing the performance of other semiclassical methods, and extending the MF method via the so-called quasi-classical formalism~\cite{Bojowald:2005cw,Bojowald:2010qm,Baytas:2018gbu,Baytas:2018ruy,Bojowald_2021,Bojowald_2022}. Lastly, an amusing side project would be to turn the logic of this paper on its head and explore the possibility of devising an interaction that classicalizes a quantum state, taking for instance a highly excited harmonic oscillator state $\ket{n}$ and naturally evolving it into a coherent state $\ket{\alpha}$ with average occupation number $|\alpha|^2=n$.

% \fd{I think the outline of the conclusions should be:
% \begin{enumerate}
%     \item MF seems to require weak collective couplings. Interestingly, this coincides with the regime where many breathers live for example.
%     \item TW is adequate in describing situations where entanglement plays a role. This is somehow tied to the classical thermalization of the classical paths of particles.
%     \item TW is essentially what is applied in preheating with an extra assumption of squeezing.
% \end{enumerate}
% Something like that but happy with anything.}

\section*{Acknowledgments}
We would like to thank Eugenio Bianchi, Martin Bojowald, Tanmay Vachaspati, and Juan Sebastien Valbuena for useful discussions. F.~D. acknowledges the support from the Departament de Recerca i Universitats from Generalitat de Catalunya
to the Grup de Recerca 00649 (Codi: 2021 SGR 00649) and funding from the ESF under the program Ayudas predoctorales of the Ministerio de Ciencia e Innovación PRE2020-094420. G.~Z. is supported by SONATA BIS Grant No. 2023/50/E/ST2/00231
from the Polish National Science Centre. This work is part of the doctoral thesis of F.~D. within the framework of the Doctoral Program in Physics at the Autonomous University of Barcelona. The data used to produce the figures in this work are available at \cite{UJ/ZWASYL_2025}.

\appendix

\section{Analytical Results for Small Collective Coupling}
\label{app:analytics}
It is instructive to understand how our toy model behaves when the coupling between the two degrees of freedom is weak. When this is the case, it can generally be expected that the ``classical'' $x$ variable is weakly entangled with the ``quantum'' $y$ variable. The first part of this appendix aims to show this explicitly. Then we will show that the MF method can exactly reproduce the average expectation value $\langle \hat{x}(t) \rangle$ for this weakly entangled state, while the Truncated Wigner method will never be able to achieve this, supporting our observations in Sec.~\ref{sec:results}. 
% We first go through the quantum mechanical calculation, followed by an analysis of the performance of the two methods MF and TW in this regime.

\subsection{Quantum Mechanics}
In this derivation, we closely follow Ref.~\cite{Vachaspati_2017}. We start by writing the quantum state of the coupled system as
\beq
\ket{\psi} = \sum_{n,m = 0}^\infty c_{n,m}(t) f_n(t) \ket{n}_x \ket{m}_y,
\label{eq:QMstatedecomp}
\eeq
where $\ket{n}_x$ and $\ket{m}_y$ are eigenstates of the free Hamiltonians for $x$ and $y$ in Eq.~\eqref{eq:dimensionlessham} respectively. The functions $f_n(t)$ are introduced for future convenience and correspond to the expansion coefficients of the initial coherent state for $x$,
\beq
f_n(t) = e^{-it/2} e^{-|z_0|^2} \frac{z_0^ne^{-int}}{\sqrt{n!}},
\label{eq:coherentstatefn}
\eeq
where $z_0$ is related to the initial displacement $x_0$ through $z_0 = x_0/\sqrt{2}$, since we assume $0$ initial momentum. All the nontrivial time dependence is thus included in the undetermined $c_{n,m}(t)$ terms. It is useful to express the position and momentum operators in terms of creation and annihilation operators via
\beqa
&x = \frac{1}{\sqrt{2}}\left(a + a^\dagger\right), \qquad p_x = \frac{i}{\sqrt{2}}\left(a^\dagger - a\right),
\label{eq:xopho}\\
&y = \sqrt{\frac{\omega}{2\Omega}}\left(b + b^\dagger\right), \qquad p_y = i \sqrt{\frac{\Omega}{2\omega}}\left(b^\dagger - b\right).
\label{eq:yopho}
\eeqa
Then the Hamiltonian \eqref{eq:dimensionlessham} reads
\beq
H = \left(a^\dagger a + \frac{1}{2}\right) + \frac{\Omega}{\omega}\left(b^\dagger b + \frac{1}{2}\right) + \frac{\lambda \omega}{2\Omega} \frac{(a + a^\dagger)^2}{2} \frac{(b + b^\dagger)^2}{2}.
\label{eq:hamiltonosc}
\eeq
Note that the effective coupling now appears to be $\lambda\omega/\Omega$, however in what follows we will assume $\Omega/\omega \sim \mathcal{O}(1)$. Plugging the ansatz~\eqref{eq:QMstatedecomp} into the Schr\"odinger equation with Hamiltonian \eqref{eq:hamiltonosc}, we obtain a coupled set of equations for the different $c_{n,m}(t)$. Until now, the steps are completely general and the set of equations we obtain in this way have no closed form. As shown in~\cite{Vachaspati_2017}, one can make progress by using the perturbative expansion
\beq
c_{n,m}(t) = \delta_{m0}e^{-i\frac{\Omega}{2 \omega}t} + \mathcal{O}(\lambda).
\eeq
This expansion implicitly assumes weak collective coupling $\lambda z_0^2 = \lambda N \ll 1$ in order to match with the setup of our problem on the one hand, and to be consistent perturbatively on the other. Without showing all the details of the computation (we refer the reader to \cite{Vachaspati_2017} for the complete version) it turns out that only the $m=0$ and $m=2$ terms get non-trivial corrections at leading order in $\lambda$, and the result can be elegantly written, in a form valid for all times, as 
\beqa
c_{n,0}(t) &=& e^{-i\frac{\Omega}{2\omega}t}\left[e^{-i\frac{\lambda \omega}{8\Omega} (2 n + 1) t } - i  \frac{\lambda \omega}{8\Omega} \left(z_0^2 e^{-it} + \frac{n(n-1)}{z_0^2}e^{it}\right)\sin t\right],
\label{eq:cn0}\\
c_{n,2}(t) &=& -i \frac{\lambda \omega}{4 \sqrt{2} \Omega}e^{-i\frac{3\Omega}{2\omega}t}\left[e^{-it} z_0^2 \frac{\sin((\Omega/\omega - 1) t)}{\Omega/\omega -1} + (2n + 1) \frac{\sin((\Omega/\omega) t)}{\Omega/\omega}\right.\nonumber\\ 
&&\hspace{2.6in}\left.+ e^{it} \frac{n(n-1)}{z_0^2} \frac{\sin((\Omega/\omega + 1) t)}{\Omega/\omega +1}\right].
\label{eq:cn2}
\eeqa
% \beq
% c_{n,0}(t) = e^{-i\frac{\Omega}{2\omega}t}\left[e^{i\frac{\lambda \omega}{\Omega} (2 n + 1) t \l} - i  \frac{\lambda \omega}{\Omega} \left(z_0^2 e^{-it} + \frac{n(n-1)}{z_0^2}e^{it}\right)\sin(t)\right]
% \label{eq:cn0}
% \eeq
% \beq
% c_{n,2}(t) = -i \frac{\lambda \omega}{4 \sqrt{2} \Omega}\left(e^{-it} z_0^2 \frac{\sin((\omega - 1) t)}{\omega -1} + (2n + 1) \frac{\sin(\omega t)}{\omega} + e^{it} \frac{n(n-1)}{z_0^2} \frac{\sin((\omega + 1) t)}{\omega +1}\right)
% \label{eq:cn2}
% \eeq
The requirement of small collective coupling now becomes evident since terms of order $\lambda z_0^2$ make their appearance. Also note that the terms inversely proportional to $z_0$ become unimportant as $N \rightarrow \infty$ but we will keep the discussion general. Of course, in this work we are not interested in the quantum mechanical state itself, but instead on observables such as $\langle \hat{x}(t) \rangle$. Using \eqref{eq:xopho} we find 
\beq
\langle \hat{x}(t) \rangle = \frac{z_0}{\sqrt{2}} \sum_{n,m = 0}^\infty \left( e^{-it} c_{n+1,m}c_{n,m}^* + e^{it} c_{n+1,m}^*c_{n,m}\right)|f_n|^2,
\eeq
and plugging in the solutions from Eqs.~\eqref{eq:cn0} and~\eqref{eq:cn2} we obtain to lowest order in the coupling
\footnote{Note that in \cite{Vachaspati_2017} the perturbative expansion was organized in such a way that corrections to the frequency and amplitude of the oscillations could be treated separately. In this way the perturbative expansion is under control for all times. The result given here is zeroth order in the amplitude and first order in the frequency.}
\beq
\langle \hat{x}(t) \rangle \approx x_0 \cos\left[t\left(1 + \lambda \frac{\omega}{4 \Omega}\right)\right].
\label{eq:qmvev}
\eeq
The behavior of higher moments is also important in order to ascertain how classical the state looks. For instance we find
\beq
\langle \hat{x}^2(t) \rangle = \sum_{n,m = 0}^\infty \left(\frac{1}{2}z_0^2 e^{-2it} c_{n+2,m}c_{n,m}^* +\frac{1}{2}z_0^2 e^{2it} c^*_{n+2,m}c_{n,m}+ \left(n +\frac{1}{2}\right) |c_{n,m}|^2\right)|f_n|^2,
\eeq
and plugging in the solutions~\eqref{eq:cn0} and~\eqref{eq:cn2} we obtain
\beq
\langle \hat{x}^2(t) \rangle \approx \frac{x_0^2}{2}\left(1+\cos\left[2 t\left(1 + \lambda \frac{\omega}{4 \Omega}\right)\right]\right) + \frac{1}{2} = \langle \hat{x}(t)\rangle^2 + \frac{1}{2}.
\label{eq:qmvar}
\eeq
This result suggests that, to leading order in $\lambda z_0^2$, time evolution preserves the approximately classical nature of the state (see Sec.~\ref{sec:motivation} and Ref.~\cite{Glauberfact}). In other words, all the information about the state of the $x$ variable is encoded in $\langle \hat{x}(t) \rangle$ suggesting that the state remains coherent. It simply transitions to a coherent state with a different frequency. In the next subsection we show that this type of behavior is mimicked perfectly by the MF method in the weak collective coupling limit.

Finally, we make a brief comment regarding entanglement. In the main part of the text, we suggest that entanglement is intimately tied to the breakdown of semiclassicality and also plays a vital role in determining which method performs better between MF and TW. Using~\eqref{eq:cn0} it is easy to see that the state $\ket{\psi}$ is a product state at leading order in $\lambda$,
\beq
\ket{\psi} = \sum_{n,m=0}^\infty \delta_{m0} e^{-i\frac{\Omega}{2\omega}t}e^{-i\frac{\lambda \omega}{8\Omega} (2 n + 1) t }f_n(t)\ket{n}_x\ket{m}_y + \mathcal{O}(\lambda),
\label{eq:entangledstate}
\eeq
and that quantum entanglement only arises at higher order in the collective coupling. In this low entanglement limit quantum effects seem to mainly manifest as a correction to the oscillation frequency of the classical solution. In what follows we will show that this is well captured by the MF method.

\subsection{Mean-Field}
The MF equations of motion are Eqs.~\eqref{eq:classCQC} and~\eqref{eq:qmCQC} and we wish to find a perturbative solution with the appropriate initial conditions. Previously, we saw that in the limit of small couplings, the main correction to the classical solution $x(t) = x_0 \cos t$ was a change in the frequency, proportional to $\lambda$. With this in mind, we perform a two-timing analysis in which we introduce an extra timescale $\tau = \frac{\lambda \omega}{\Omega} t$. Then the perturbative solution can be obtained via the replacement
\beq
\partial_t^2 \rightarrow \partial_t^2 + 2 \frac{\lambda \omega}{\Omega} \partial_t\partial_\tau + \mathcal{O}(\lambda^2)
\eeq
as well as a standard expansion of the solutions,
\beqa
x &=& x^{(0)} + \lambda x^{(1)} + \mathcal{O}(\lambda^2),\\
z &=& z^{(0)} + \lambda z^{(1)} +\mathcal{O}(\lambda^2).
\eeqa
Plugging this into Eqs.~\eqref{eq:classCQC} and~\eqref{eq:qmCQC} it is straightforward to find the general solutions for $x^{(0)}$ and $z^{(0)}$. They are given by
\beqa
x^{(0)} &=& x_0 \cos(t + \delta_x(\tau)),\\
z^{(0)} &=& -i \sqrt{\frac{\omega}{2 \Omega}} e^{i ((\Omega/\omega) t + \delta_z(\tau))},
\eeqa
where we allow for a dependence on $\tau$ of the phases. Without loss of generality we may set $\delta_x(\tau)$ and $\delta_z(\tau)$ to 0 at $t = \tau = 0$. We can then use the first-order equations of motion to find the form of these phase corrections. Since we are mainly interested in the behavior of $x$ we only write down the zeroth-order equation for $x^{(1)}$ and we again assume small collective coupling $\lambda x_0^2 \ll 1$. We find
\beq
\Ddot{x}^{(1)} + x^{(1)} = -\frac{2\omega}{\Omega}\partial_t \partial_\tau x^{(0)} - |z^{(0)}|^2 x^{(0)}.
\eeq
After plugging in the zeroth order solutions we obtain
\beq
\Ddot{x}^{(1)} + x^{(1)} = \left(\frac{2 \omega}{\Omega} \partial_\tau \delta_x(\tau) - \frac{\omega}{2 \Omega}\right) x_0 \cos(t + \delta_x(\tau)).
\label{eq:firstorderCQC}
\eeq
If the right hand side of this equation is not $0$, the solution for $x^{(1)}$ would resonate, resulting in an immediate breakdown of perturbation theory. This yields the condition
\beq
\delta_x(\tau) = \frac{1}{4} \tau = \frac{\lambda \omega}{4 \Omega} t.
\eeq
In principle, we can continue this procedure to higher orders, but at this point we have perfect agreement between the MF solution,
\beq
x(t) \approx x^{(0)}(t) = x_0 \cos\left[\left(1 + \lambda \frac{\omega}{4 \Omega}\right)t\right],
\eeq
and the quantum mechanical result~\eqref{eq:qmvev}. It is remarkable that in the weak collective coupling regime the MF method can reproduce quantum mechanical results for very long times. What's more, since the average expectation value completely determines the quantum mechanical state in this regime (see~\eqref{eq:qmvar}), the MF method manages to reproduce the results of a full quantum mechanical simulation at just a fraction of the computational cost. We will now end by arguing that this desirable property is not shared by the TW method.

\subsection{Trunctated Wigner}
It is significantly more difficult to derive an analytic estimate for the results of the TW method. However, here we will argue that after some time it will necessarily deviate from the weak collective coupling behavior of quantum mechanics. The best way to see this is to inspect one of the many classical realizations obtained in the TW framework. Applying the two-timing perturbative expansion of the previous section to the classical equations of motion~\eqref{eq:class} and~\eqref{eq:quantum}, it is straightforward to show that the solution for the ``classical'' $x$ variable approximately reads
\beq
x(t) \approx x_0 \cos\left[\left(1 + \lambda \frac{\omega}{4 \Omega}\left( (\Omega/\omega) y_0^2 + (\omega/\Omega) p_{y,0}^2\right)\right)t\right],
\eeq
for a particular, randomly sampled, initial condition for the ``quantum'' $y$ variable, $(y_0,p_{y,0})$. Looking at this expression we see that the phases of the various classical realizations become completely uncorrelated within a time of order $\lambda^{-1}\left(\langle y_0^2\rangle + (\omega/\Omega)^2 \langle p_{y,0}^2\rangle\right)^{-1}\sim \lambda^{-1}(2\omega/\Omega)^{-1}$ (where the brackets denote a statistical average weighted by the distribution~\eqref{eq:Wignerinit}). After this time it should be approximately equivalent to write each solution as
\beq
x(t) \approx x_0 \cos\left(t + \phi\right),
\label{eq:twsollatetime}
\eeq
where $\phi$ is some random phase. At large times, the TW method thus seems to predict that the $x$ variable slowly dissipates, as averaging over an ensemble of solutions such as the ones in \eqref{eq:twsollatetime} will inevitably lead to amplitude decay. This behavior is not the one predicted by quantum mechanics in a classically stable region of parameter space, where the system settles to a new equilibrium solution and there is no dissipation. These results explain why the MF method performs better than TW in regions where the collective coupling is small. The fact that the TW method seems to explore the full phase space of the theory, which helps the method in regions of classical instability, hurts it when quantum entanglement doesn't grow uncontrollably.  

\section{Systems with more degrees of freedom}
\label{app:threemodes}

This appendix constitutes a preliminary investigation of how our results extend to systems with more degrees of freedom. In field theory, there are many modes, of which some can be stable, while others can be unstable. To understand how our results extend to quantum field theory it is thus important to introduce more degrees of freedom in our analysis. 

As a first step, we analyze a three mode system defined through the dimensionless Hamiltonian
\beq
H = -\frac{1}{2}\partial^2_x+ \frac{1}{2} x^2 -\frac{1}{2}\partial^2_y+ \frac{1}{2} \left[\left(\Omega_1/\omega\right)^2 + \lambda x^2\right] y^2 -\frac{1}{2}\partial^2_z+ \frac{1}{2} \left[\left(\Omega_2/\omega\right)^2 + \lambda x^2\right] z^2.
\label{eq:dimensionlesshamthreemode}
\eeq
Here both the $y$ and $z$ degree of freedom should be interpreted as quantum modes, while $x$ still corresponds to the classical background. We perform the computations in the same way as for the two-mode system, focusing on the break times of the different methods as well as on the entanglement entropy of $x$ (this time the density matrix is partially traced over both the $y$ and $z$ degrees of freedom whose Hilbert space has been restricted to a finite dimensional subspace). We are particularly interested in cases where: (i) both modes are classically stable, (ii) both modes are classically unstable, and (iii) one mode is unstable while the other is stable. Results are shown in Table~\ref{tab:twomode}.

\begin{table}[h!]
\centering
\begin{tabular}{|c|c|c|c|c|c|c|c|c|}
\hline
$\lambda$ & $x_0$& $\Omega_1/\omega$&  $\Omega_2/\omega$ & Stability & $t_q$ & $t_{MF}$& $t_{TW}$& $S_e$ at $t = 25$\\ \hline
0.1      & 5      & 0.4      & 0.8      & UU      & 4      & 9      & 17& 2.93     \\ \hline
0.05      & 8      & 0.4      & 0.8      & US      & 7      & 12      & 25& 2.37     \\ \hline
0.04    & 6     & 0.8     & 1.2    & US      & 11      & 19      & 31      & 1.62      \\ \hline
0.03     & 5     & 0.4     & 1.2     & SS     & 11      &83     & 70     & 0.34     \\ \hline
\end{tabular}
\caption{Some prototypical results for the three mode system. Three situations can be distinguished: UU, when both modes are unstable, US, when the first mode is unstable while the second one is stable, and SS when both modes are stable. We can readily compare these results to those of Table~\ref{tab:singlemode}, for the two mode system. Again the influence of the existence of unstable modes  on the break times is evident.}
\label{tab:twomode}
\end{table}
Our conclusions about the influence of classical instability and entanglement on the break times seem to extend to the three mode system. Comparing the above results to the analogue ones for the two mode case (see~Table~\ref{tab:singlemode}), we see that the ``most entangling'' mode ultimately determines the entanglement entropy and break times of the three mode system. However the amount of unstable modes also seems to play a role. It is in particular unclear to what extent the presence of more stable modes can wash out the effect of the instability and extend the validity of semiclassical methods. Unfortunately incorporating more modes quickly becomes computationally expensive, and is thus beyond the scope of this work.

\section{Numerics and performance tests}
\label{app:numerics}
In this appendix, we expand on the numerical integration schemes that we employed in this work. We had to solve three different sets of equations corresponding to the various methods under investigation. To be precise, the Mean Field equations~\eqref{eq:qmCQC} and \eqref{eq:classCQC}, the Truncated Wigner equations (which are just the classical equations of motion~\eqref{eq:class} and~\eqref{eq:quantum}, evolved for a large number of different initial conditions), and the full Schr\"odinger equation of quantum mechanics. Since the Schr\"odinger equation is a wave equation requiring lattice discretization to solve, we will mostly focus on the latter. As for the MF and TW methods, we will just mention that we used a \textit{VV8} integration scheme for most of the results shown in this paper \cite{YOSHIDA1990262}. For certain prototypical points in parameter space (in particular in the corners of Figs.~\ref{fig:w0d4grid},~\ref{fig:w0d8grid} and~\ref{fig:w1d2grid}) we explicitly checked that our results were stable against decreasing the time increment $\Delta t$. It is also good to mention that numerical results in the TW method were obtained by averaging over $N_s=8192$ different realizations. Again, we checked for some prototypical parameter values that a higher number of samples did not change the results.

Now we turn our attention to the integration scheme used for the Schr\"odinger equation
\beq
i \frac{\partial\psi}{\partial t} = H \psi,
\eeq
where the Hamiltonian is written in position space as in Eq.~\eqref{eq:dimensionlessham}. The initial conditions given in Eq.~\eqref{eq:wavefunctioninit} completely determine the time evolution of the system. Formally this is given by
\beq
\psi(t,x,y) = e^{-i \int_0^t H dt} \psi_0(x,y).
\label{eq:timeint}
\eeq
To find a numerical approximation of this solution we need to discretize space on a grid, writing $x_j = j \Delta x$ and $y_j = j \Delta x$, and solve the equation in a finite domain defined to be a square box of size $L$ (centered around the origin). The number $N^2$ of gridpoints we include determines the spatial resolution $\Delta x = L/N$. Finally, we also discretize time, by defining $t_j = j \Delta t$, and evolve the wavefunction by rewriting~\eqref{eq:timeint} as
\beq
\psi(t_f,x_j, y_j) = \Pi_{s=0}^f e^{-i\frac{\Delta t}{2} H_k}  e^{-i\Delta t H_p}  e^{-i\frac{\Delta t}{2} H_k} \psi_0(t_0, x_j, y_j),
\eeq
where $H_k$ denotes the kinetic part of the Hamiltonian (the spatial derivatives) and $H_p$, the potential part. In practice, we employ a pseudo-spectral method where we evolve with $H_k$ in Fourier space by rotating the discrete Fourier transform (DFT) of $\psi(x_i,y_i, t_i)$ twice at each timestep. Notice that this is simply a \textit{VV2} integration scheme applied to the discretized Schr\"odinger equation \cite{Levkov_2018}. It is a symplectic integrator, conserving the total probability and average energy to high precision. Our method implicitly assumes periodic boundary conditions due to the way the DFT is performed. We checked that the boundary has little effect on our numerical solution, as long as the initial coherent state is well separated from it. This is because we are always working with a confining potential. To check that our results are not too sensitive to the spatial and temporal resolutions ($\Delta x$ and $\Delta t$), and to the finite size of the box ($L$) we have performed convergence tests of the code, and investigated the influence of each source of truncation error individually. We did this by defining benchmark simulation parameters, $N = 256$, $L = 40$, $\Delta x = L/N \approx 0.15$ and $\Delta t = 0.03$, and comparing the resulting numerical solution to the one obtained by changing either $\Delta t$, $\Delta x$, or $L$, keeping everything else constant. Note that the benchmark parameters are precisely the ones we used for most of the results that we present in the main body of the text. 

\begin{figure}[h!]
    \centering
    \includegraphics[width=0.8\textwidth]{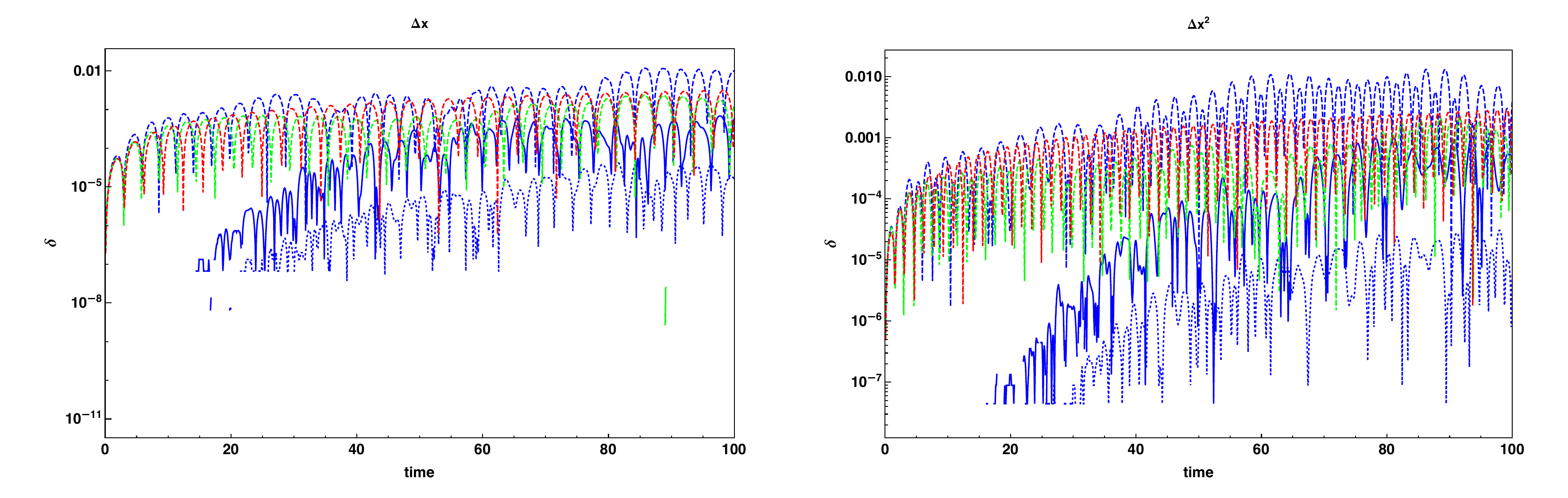}
    \caption{Plot of the convergence parameter as defined in Eq.~\eqref{eq:convergenceparam} as a function of time, for the operators $\hat{x}$ (left) and $\hat{x}^2$ (right). The results are shown for three representative points in our physical parameter space, namely $\Omega/\omega = 0.8, \lambda = 0.1, x_0 = 15$ (blue), $\Omega/\omega = 0.4, \lambda = 0.1, x_0 = 4$ (green) and $\Omega/\omega = 1.2, \lambda = 0.04, x_0 = 10$ (red). We vary the benchmark simulation parameters in three different ways: reducing $\Delta t$ by a factor of two (dashed line), reducing $\Delta x$ by a factor of two (dotted line), and increasing the box size $L$ by a factor of two (continuous line). In some cases the differences were smaller than machine precision, which is why they were not plotted here. The influence of $\Delta t$ is the most important although it is limited. Note that in the case where the error becomes $\mathcal{O}(10^{-2})$ (dashed blue line), the associated break times are much shorter (see Fig.~\ref{fig:w0d8grid}) than the 100 time units that are shown here and therefore our results can still be trusted.}
    \label{fig:convergence}
\end{figure}

For example, we can compare how the average value of some operator $\hat{O}$, $\langle \hat{O}(t) \rangle$, changes when varying one of the benchmark parameters, by plotting the quantity\footnote{Note that the quantity $\delta$ introduced here has nothing to do with the one in in Eq.~\eqref{eq:delta2}.}
\beq
\delta(t) = \frac{|\langle \hat{O}(t) \rangle - \langle \hat{O}(t) \rangle_b|}{\langle \hat{O}(t=0) \rangle},
\label{eq:convergenceparam}
\eeq
where the subscript $b$ refers to the benchmark simulation and the normalization factor in the denominator is exactly computable in terms of the initial conditions. In Fig.~\ref{fig:convergence}, we show the result of this test for operators $\hat{x}$ and $\hat{x}^2$ for three representative points in physical parameter space.

We observe that the errors remain small for all times. In fact, the maximum error is obtained when varying $\Delta t$. For $\Omega/\omega = 0.8$, $\lambda = 0.1$ and $x_0 = 15$ it becomes  of order $10^{-2}$. These errors do not influence the conclusions of this work as they are too small to change the results in any meaningful way. Of course, we could have reduced the error of our numerics by increasing the resolution of our grid, but opted not to in order to keep the computational time within reasonable limits. We believe that the results of Fig.~\ref{fig:convergence} warrant this decision.

\begin{figure}
    \centering
    \includegraphics[width=0.7\textwidth]{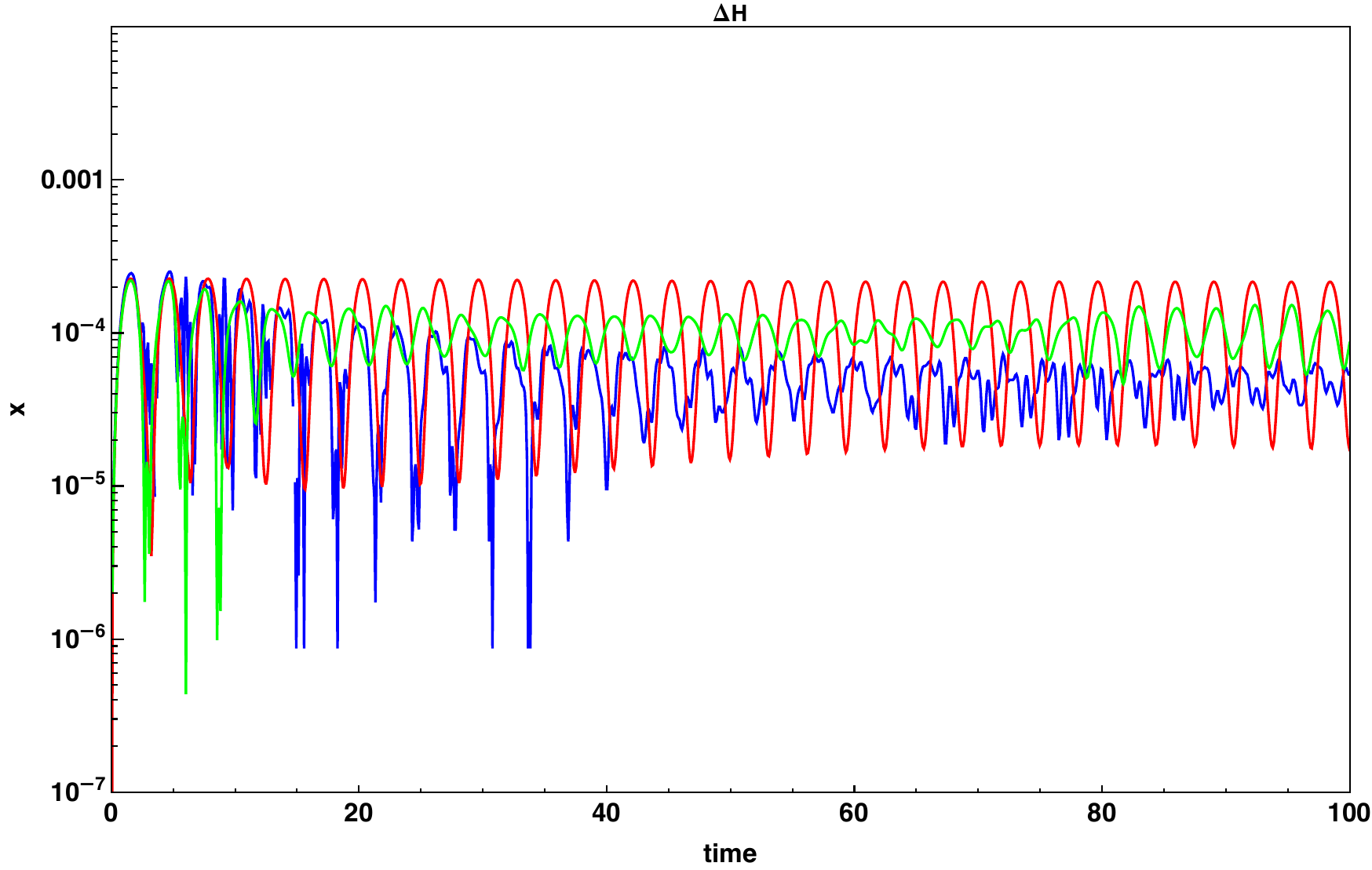}
    \caption{The degree of conservation of the expectation value of $H$ for the benchmark simulations of the previous three choices of physical parameters. In all cases energy is conserved up to 1 part in $10^4$.}
    \label{fig:hconserve}
\end{figure}

As a final test of our numerics, we also explicitly show that the average energy is conserved (by plotting $\Delta H(t)=|1 - \langle H(t) \rangle/\langle H(t=0) \rangle$|)for our benchmark simulation parameters. This is shown in Fig.~\ref{fig:hconserve}. 
As expected we see that our code conserves energy up to 1 part in $10^4$.

\newpage

\bibliography{CQCVsS}

\end{document}